\documentclass[12pt]{article}
\usepackage{epsfig}\parskip 5pt plus 1pt
\usepackage{bbm}
\usepackage{amsmath}
\usepackage{amssymb}
\usepackage{amsfonts}
\usepackage{mathrsfs}
\usepackage{graphicx}
\usepackage{fancybox}
\usepackage{appendix}
\usepackage{pifont}
\usepackage{color}
\newcommand{\nn}{\nonumber}

\newcommand{\dv}{\partial\hspace{-7pt}\slash}

\newcommand{\be}{\begin{equation}}
\newcommand{\ee}{\end{equation}}
\newcommand{\bea}{\begin{eqnarray}}
\newcommand{\eea}{\end{eqnarray}}

\newcommand{\Ve}{V_e}
\newcommand{\Vm}{V_{\mu}}
\newcommand{\Vt}{V_\tau}

\newcommand{\U}{(\mathcal{U}_{PMNS})}

\begin{document}

\begin{titlepage}
\vspace*{-2cm}
\hfill{UAB-FT-694}
\vskip 3cm
\begin{center}
{\Large\bf Implementation of the type III seesaw model in
  FeynRules/MadGraph and prospects for discovery with early LHC data}
\end{center}
\vskip 0.5cm
\begin{center}
{\large C.~Biggio}$\,^a$~\footnote{biggio@ifae.es} and
{\large F.~Bonnet}$\,^b$~\footnote{florian.bonnet@pd.infn.it}\\
\vskip .7cm
$^a\,$ Institut de F\'\i sica d'Altes Energies, Universitat Aut\`onoma de Barcelona, 08193 Bellaterra, Spain
\vskip .1cm
$^b\,$ Istituto Nazionale di Fisica Nucleare, Sezione di Padova, via Marzolo 8, 35131 Padova, Italy
\end{center}
\vskip 0.5cm

\begin{abstract}
We discuss the implementation of the ``minimal'' type III seesaw
model, i.e. with one fermionic triplet, in FeynRules/MadGraph. This is
the first step in order to realize a real study of LHC data recorded
in the LHC detectors. With this goal in mind, we comment on the
possibility of discovering this kind of new physics at the LHC running
at 7~TeV with a luminosity of few fb$^{-1}$.
\end{abstract}
\end{titlepage}
\setcounter{footnote}{0}
\vskip2truecm


\section{Introduction}

In a period in which LHC is running and ready to discover new physics,
it is of crucial importance to have the possibility of simulating the
signals that a particular kind of new physics could give in the two
main detectors, ATLAS and CMS. In this paper we describe the
implementation in
\texttt{FeynRules/MadGraph}~\cite{Christensen:2008py,MadGraph} of
a simple extension of the standard model (SM), the ``minimal'' type
III seesaw. This is a first necessary step before performing the
analysis of real data, which is the ultimate goal of our work and
which will be discussed in a future publication.

As it is well known, oscillation experiments have proved that
neutrinos oscillate and therefore are massive. However, from the
theoretical point of view, the origin of this mass is still
unknown. An appealing possibilty, also accounting for the smallness of
this mass, is the seesaw mechanism: new heavy states having a Yukawa
interaction with the lepton and the Higgs doublets generate a small
Majorana mass for the neutrinos, generically suppressed, with respect
to charged fermion masses, by a factor $v/M$, where $v$ is the Higgs
vev and $M$ the mass of the heavy particle. Depending on the nature of
the heavy state, seesaw models are called type I~\cite{TypeI}, type
II~\cite{TypeII} or type III~\cite{TypeIII}, corresponding to heavy fermionic
singlet, scalar triplet or fermionic triplet, respectively. If one
requires $\mathcal{O}(1)$ Yukawa couplings, $M$ should be of the order
of the grand unification scale in order to account for neutrino masses
smaller than the eV. However, in principle the scale can be as low as
hundreds of GeV, in which case either the Yukawas are smaller or an
alternative method, such as for instance an inverse seesaw~\cite{ISS}
should be at work. In this case the heavy field responsible for
neutrino masses could be discovered at the LHC.

As regards collider physics, the seesaws of type II and III are more
exciting, since they can be produced via gauge interactions: at
difference with singlets, whose production is drastically suppressed
if the Yukawa couplings are small, triplets can be produced and
observed at the LHC if their mass is sufficiently small, independently
of the size of the Yukawa couplings or mixing angles.

In the present paper we focus on the type III seesaw, i.e. the one
mediated by fermionic triplets. To simplify the implementation of the
model in \texttt{FeynRules}, we consider a simple extension of the SM
obtained by adding a single triplet. Indeed we can safely assume that,
unless in case of extreme degeneracy, the lightest triplet will be the
one most copiously produced and the one which will be eventually
firstly discovered. In the literature few
papers~\cite{delAguila:2008cj,Franceschini:2008pz, Senjanovic}
discussing the possibility of discovering the type III seesaw at the
LHC (at 14~TeV) are present. However so far no code is publicly
available to perform calculations and simulations in this model. With
this paper and the publication of the implemented model at the URL
\texttt{http://feynrules.phys.ucl.ac.be/wiki/TypeIIISeeSaw} we are
going to fill this gap. Moreover we briefly discuss the physics case
for LHC running at 7~TeV, suggesting that with few fb$^{-1}$ of
luminosity a discovery is already possible.

This paper is organized as follows. In Sect.~2 the model with the
complete Lagrangian and all the couplings is reviewed, both in the
general and in the simplified case. In Sect.~3 the implementation of
the model in \texttt{FeynRules} and the checks performed for its
validation are discussed. In Sect.~4 the physics case at 7~TeV is
discussed and in Sect.~5 we conclude.


\section{The model}

The model considered here is the one presented in
Ref.~\cite{Abada:2008ea}. It consists in the addition to the standard
model of SU(2) triplets of fermions with zero hypercharge, $\Sigma$.
In this model at least two such triplets are necessary in order to
have two non-vanishing neutrino masses. The beyond the standard model
interactions are described by the following lagrangian (with implicit
flavour summation):
\begin{equation}
\label{Lfermtriptwobytwo}
{\cal L}=Tr [ \overline{\Sigma} i \slash \hspace{-2.5mm} D  \Sigma ] 
-\frac{1}{2} Tr [\overline{\Sigma}  M_\Sigma \Sigma^c 
                +\overline{\Sigma^c} M_\Sigma^* \Sigma] 
- \tilde{\phi}^\dagger \overline{\Sigma} \sqrt{2}Y_\Sigma L 
-  \overline{L}\sqrt{2} {Y_\Sigma}^\dagger  \Sigma \tilde{\phi}\, ,
\end{equation} 
with $L\equiv (\nu , l)^T$, $\phi\equiv (\phi^+, \phi^0)^T\equiv
(\phi^+, (v+H+i \eta)/\sqrt{2})^T$, $\tilde \phi = i \tau_{2} \phi^*$,
$\Sigma^c \equiv C \overline{\Sigma}^T$ and with, for each fermionic
triplet,
\begin{eqnarray}
\Sigma&=&
\left(
\begin{array}{ cc}
   \Sigma^0/\sqrt{2}  &   \Sigma^+ \\
     \Sigma^- &  -\Sigma^0/\sqrt{2} 
\end{array}
\right), \quad 
\Sigma^c=
\left(
\begin{array}{ cc}
   \Sigma^{0c}/\sqrt{2}  &   \Sigma^{-c} \\
     \Sigma^{+c} &  -\Sigma^{0c}/\sqrt{2} 
\end{array}
\right), \nonumber\\
D_\mu &=& \partial_\mu-i\sqrt{2} g \left(
\begin{array}{ cc}
   W^3_\mu/\sqrt{2}  &   W_\mu^+ \\
     W_\mu^- &  -W^3_\mu/\sqrt{2} 
\end{array}
\right)\,.
\end{eqnarray}
Without loss of generality, we can assume that we start from the basis
where $M_\Sigma$ is real and diagonal, as well as the charged lepton
Yukawa coupling, not explicitly written above.  In order to consider
the mixing of the triplets with the charged leptons, it is convenient
to express the four degrees of freedom of each charged triplet in
terms of a single Dirac spinor:
\begin{equation}
 \label{Psi}
\Psi\equiv\Sigma_R^{+ c} + \Sigma_R^-\,.
\end{equation}
The neutral fermionic triplet components on the other hand can be left
in two-component notation, since they have only two degrees of
freedom and mix with neutrinos, which are also described by
two-component fields. This leads to the Lagrangian
\begin{eqnarray}
\label{Lfull-ft-2}
{\cal L}&=& \overline{\Psi} i \dv \Psi  
+ \overline{\Sigma_R^0} i \dv  \Sigma^0_R  
-  \overline{\Psi}M_\Sigma \Psi -
        \left( \overline{\Sigma^{0}_R} \frac{{M_\Sigma}}{2}  \Sigma_R^{0c} \,+  \,\text{h.c.}\right) 
\nonumber \\
&+&g \left(W_\mu^+ \overline{\Sigma_R^0} \gamma_\mu  P_R\Psi 
 +  W_\mu^+ \overline{\Sigma_R^{0c}} \gamma_\mu  P_L\Psi   \,+  \,\text{h.c.}
 \right) - g\, W_\mu^3 \overline{\Psi} \gamma_\mu  \Psi 
 \nonumber\\
\nonumber \\ 
&-&  \left( \phi^0 \overline{\Sigma_R^0} Y_\Sigma \nu_{L}+ \sqrt{2}\phi^0 \overline{\Psi} Y_\Sigma l_{L}
+     \phi^+ \overline{\Sigma_R ^0} Y_\Sigma l_{L} - \sqrt{2}\phi^+
 \overline{{\nu_{L}}^c}
Y^{T}_\Sigma \Psi   \,+  \,\text{h.c.}\right)\,.
\end{eqnarray}
The mass matrices of the charged and the neutral sectors need to be
diagonalized as they possess off-diagonal terms.  Following the
diagonalization procedure described in Ref.~\cite{Abada:2008ea}, we
obtain the following Lagrangian in the mass basis:
\begin{equation}
\mathcal{L}=\mathcal{L}_{Kin}+\mathcal{L}_{CC}+\mathcal{L}_{NC}^{\ell}+\mathcal{L}_{NC}^{\nu}+\mathcal{L}_{H}^{\ell}+\mathcal{L}_{H}^{\nu}+\mathcal{L}_{\eta}^{\ell}+\mathcal{L}_{\eta}^{\nu}+\mathcal{L}_{\phi^-}\, ,
\label{fulllagrangian}
\end{equation}
where
%
%
\bea
\label{CC}
\mathcal{L}_{CC}&=&\frac{g}{\sqrt{2}}\left(\begin{array}{cc}\overline{l} & \overline{\Psi}\end{array}\right)\gamma^{\mu}W^{-}_{\mu}\left(P_L g^{CC}_L+P_R g^{CC}_R\sqrt{2}\right)\left(\begin{array}{c}\nu \\ \Sigma\end{array}\right)+\textrm{h.c}.
\eea

\bea
\label{NCl}
\mathcal{L}_{NC}^{\ell}&=&\frac{g}{cos\theta_W}\left(\begin{array}{cc}\overline{l} & \overline{\Psi}\end{array}\right)\gamma^{\mu}Z_{\mu}\left(P_L g^{NC}_L+P_R g^{NC}_R\right)\left(\begin{array}{c}l \\ \Psi\end{array}\right)\\
%
\label{NCn}
\mathcal{L}_{NC}^{\nu}&=&\frac{g}{2cos\theta_W}\left(\begin{array}{cc}\overline{\nu} & \overline{\Sigma^{0c}}\end{array}\right)\gamma^{\mu}Z_{\mu}\left(P_L g^{NC}_{\nu}\right)\left(\begin{array}{c}\nu_L \\ \Sigma^{0c}\end{array}\right)\\
%
\label{Hl}
\mathcal{L}_{H}^{\ell}&=&-\left(\begin{array}{cc}\overline{l} & \overline{\Psi}\end{array}\right)H\left(P_L g^{H\ell}_L+P_R g^{H\ell}_R\right)\left(\begin{array}{c}l \\ \Psi\end{array}\right)\\
%
\mathcal{L}_{H}^{\nu}&=&-
\left(\begin{array}{cc}\overline{\nu} & \overline{\Sigma^0}\end{array}\right)
\frac{H}{\sqrt{2}}\left(P_L g^{H\nu}_L+P_R g^{H\nu}_R\right)
\left(\begin{array}{c}\nu \\ \Sigma^0\end{array}\right)\\
%
\label{etal}
\mathcal{L}_{\eta}^{\ell}&=&-
\left(\begin{array}{cc}\overline{l} & \overline{\Psi}\end{array}\right)
i\eta\left(P_L g^{\eta\ell}_L+P_R g^{\eta\ell}_R\right)
\left(\begin{array}{c}l \\ \Psi\end{array}\right)\\
%
\mathcal{L}_{\eta}^{\nu}&=&-
\left(\begin{array}{cc}\overline{\nu} & \overline{\Sigma^0}\end{array}\right)
\frac{i\eta}{\sqrt{2}}\left(P_L g^{\eta\nu}_L+P_R g^{\eta\nu}_R\right)
\left(\begin{array}{c}\nu \\ \Sigma^0\end{array}\right)\\
%
\mathcal{L}_{\phi^-}&=&-
\left(\begin{array}{cc}\overline{l}&\overline{\psi}\end{array}\right)
\phi^-(P_L g_L^{\phi^-}+P_R g_R^{\phi^-})
\left(\begin{array}{c}\nu\\\Sigma^0\end{array}\right)+\textrm{h.c.}
\eea
with
%
%
\bea
\label{gccL}
g^{CC}_{L}&=& 
\left(\begin{array}{cc}
\left(1+\frac{\epsilon}{2}\right)U_{PMNS} & 
            -Y_{\Sigma}^{\dagger}M_{\Sigma}^{-1}\frac{v}{\sqrt{2}}\\
0 & \sqrt{2}\left(1-\frac{\epsilon'}{2}\right)\end{array}\right)\\
\label{gccR}
g^{CC}_{R}&= &
\left(\begin{array}{cc}
0 & -m_lY_{\Sigma}^{\dagger}M_{\Sigma}^{-2}v\\
-M_{\Sigma}^{-1}Y^{*}_{\Sigma}U^{*}_{PMNS}\frac{v}{\sqrt{2}} & 
             1-\frac{\epsilon'^{*}}{2}\end{array}\right) \\
\label{gncL}
g^{NC}_{L}&=&
\left(\begin{array}{cc}
\frac{1}{2}-cos^2\theta_W-\epsilon & 
      \frac{1}{2}Y_{\Sigma}^{\dagger}M_{\Sigma}^{-1}v\\
\frac{1}{2}M_{\Sigma}^{-1}Y_{\Sigma}v & 
      \epsilon'-cos^2\theta_W\end{array}\right)\\
\label{gncR}
g^{NC}_{R}&=&
\left(\begin{array}{cc}
1-cos^2\theta_W & m_lY_{\Sigma}^{\dagger}M_{\Sigma}^{-2}v\\
 M_{\Sigma}^{-2}Y_{\Sigma}m_lv & -cos^2\theta_W \end{array}\right)\\
 \label{gncnu}
g^{NC}_{\nu}&=&
\left(\begin{array}{cc}
1-\mathcal{U}_{PMNS}^{\dagger}\,\epsilon\,\mathcal{U}_{PMNS}
& \mathcal{U}_{PMNS}^{\dagger}Y^{\dagger}_{\Sigma}M^{-1}_{\Sigma}\frac{v}{\sqrt{2}}\\
\frac{v}{\sqrt{2}}M^{-1}_{\Sigma}Y_{\Sigma}\,\mathcal{U}_{PMNS}
& \epsilon'
\end{array}\right)\\
\label{ghlL}
g^{H\ell}_{L}&=&
\left(\begin{array}{cc}
\frac{m_l}{v}\left(1-3\epsilon\right) & m_lY_{\Sigma}^{\dagger}M_{\Sigma}^{-1}\\
Y_{\Sigma}\left(1-\epsilon\right)+M_{\Sigma}^{-2}Y_{\Sigma}m_l^2 & Y_{\Sigma}Y_{\Sigma}^{\dagger}M_{\Sigma}^{-1}v \end{array}\right)\\
\label{ghlR}
g^{H\ell}_{R}&=&\left( g^{H\ell}_{L}\right)^{\dagger}\\
\label{ghnuL}
g^{H\nu}_{L}&=&
\left(\begin{array}{cc}
-\frac{\sqrt{2}}{v}\,\mathcal{U}_{PMNS}^{T}\, m_{\nu}\,\mathcal{U}_{PMNS} & \mathcal{U}_{PMNS}^T\,m_{\nu}Y_{\Sigma}^{\dagger}M_{\Sigma}^{-1} \\
(Y_{\Sigma}-Y_{\Sigma}\frac{\epsilon}{2}-\frac{\epsilon'^{T}}{2}Y_{\Sigma})\mathcal{U}_{PMNS} & Y_{\Sigma}Y_{\Sigma}^{\dagger}M_{\Sigma}^{-1}\frac{v}{\sqrt{2}} \end{array}\right)\\
&=&\left(\begin{array}{cc}
-\frac{\sqrt{2}}{v}\, m^d_{\nu} &  m^d_{\nu}\, \mathcal{U}_{PMNS}^\dagger Y_{\Sigma}^{\dagger}M_{\Sigma}^{-1} \\
(Y_{\Sigma}-Y_{\Sigma}\frac{\epsilon}{2}-\frac{\epsilon'^{T}}{2}Y_{\Sigma})\mathcal{U}_{PMNS} & Y_{\Sigma}Y_{\Sigma}^{\dagger}M_{\Sigma}^{-1}\frac{v}{\sqrt{2}} \end{array}\right)\nonumber\\
\label{ghnuR}
g^{H\nu}_{R}&=&\left( g^{H\nu}_{L} \right)^{\dagger}
\eea
\bea
\label{getalL}
g^{\eta\ell}_{L}&=&
\left(\begin{array}{cc}-\frac{m_l}{v}(1+\epsilon) & -m_l Y_{\Sigma}^{\dagger} M_{\Sigma}^{-1}\\
Y_{\Sigma}(1-\epsilon)-M_{\Sigma}^{-2}Y_\Sigma m_l^2 & v  Y_{\Sigma} Y_{\Sigma}^{\dagger}M_{\Sigma}^{-1} \end{array}\right)\\
\label{getalR}
g^{\eta\ell}_{R}&=&-\left( g^{\eta\ell}_{L}\right)^{\dagger}\\
\label{getanuL}
g^{\eta\nu}_{L}&=& g^{H\nu}_{L}\\
\label{getanuR}
g^{\eta\nu}_{R}&=&-\left( g^{\eta\nu}_{L}\right)^{\dagger}\\
\label{gphiL}
g_L^{\phi^-}&=&\left(\begin{array}{cc} 
\sqrt{2}\frac{m_l}{v}(1-\frac{\epsilon}{2})\mathcal{U}_{PMNS}
& m_l Y_\Sigma^{\dagger} M_\Sigma^{-1}\\
\sqrt{2} m_l^2 M_\Sigma^{-2} Y_\Sigma \mathcal{U}_{PMNS} 
& 0\end{array}\right)\\
\label{gphiR}
g_R^{\phi^-}&=&\left(\begin{array}{cc} 
-\sqrt{2}\mathcal{U}_{PMNS}\frac{m_{\nu}^{d *}}{v} 
& \left[(Y_\Sigma^{\dagger}-\epsilon Y_\Sigma^{\dagger}
-Y_\Sigma^{\dagger}\frac{\epsilon^{\prime *}}{2})
-2 m_{\nu}^* Y_\Sigma^\dagger M_\Sigma^{-1}\right]\\
-\sqrt{2}Y_\Sigma^*(1-\frac{\epsilon^*}{2})\mathcal{U}^*_{PMNS}
& 2[-\frac{M_\Sigma}{v}\epsilon^{\prime T} + \epsilon^\prime \frac{M_\Sigma}{v}]\end{array}\right).
\eea
Here $\mathcal{U}_{PMNS}$ is the lowest order leptonic mixing matrix
which is unitary, $m_l$ is a diagonal matrix whose elements are the
masses of the charged leptons, $v\equiv\sqrt{2}\langle \phi^0\rangle=
246$~GeV,
$\epsilon=\frac{v^2}{2}Y_{\Sigma}^{\dagger}M_{\Sigma}^{-2}Y_{\Sigma}$,
$\epsilon'=\frac{v^2}{2}M_{\Sigma}^{-1}Y_{\Sigma}Y_{\Sigma}^{\dagger}M_{\Sigma}^{-1}$and
$\delta=\frac{m_l^2}{M_\Sigma^2}$. The above expressions are all valid
at $\mathcal{O}\left(\epsilon, \epsilon^\prime, \delta,
\sqrt{\epsilon\delta}, \sqrt{\epsilon^\prime\delta} \right)$.


\subsection{The simplified model}

In the previous section the Lagrangian of the type III seesaw model,
with a generic number of triplets, has been introduced. Since we are
interested in LHC physics, we can safely restrict ourselves to the
case of only one triplet. Indeed, in the presence of more triplets, it
will be the lightest the one that will be more easily discovered. This
will simplify the implementation of the model in
\texttt{FeynRules}~\footnote{Notice that while such a simplified model
  is appropriate for studies at collider, it accounts only for one
  neutrino mass and therefore does not reproduces the experimental
  results on neutrino masses. This model should be completed with
  other heavy fields in order to obtain at
  least two massive light neutrinos. Then this simplified model should
  be viewed as a ``low''-energy limit of a more complete theory with
  heavier states that decouple. If such a hierarchy in the masses of
  the heavy particles is not realized, i.e. if, for example, two or
  more triplets are degenerate, then the analisis will be
  different. The production cross section for each of the triplet will
  be the current one, but decays would be different, due to the larger
  number of possibilities for the couplings.}.  Under this assumption, the new
Yukawa couplings matrix reduces to a $1\times 3$ vector:
\begin{equation}
Y_{\Sigma}=\left(\begin{array}{ccc} Y_{\Sigma_e} & Y_{\Sigma_\mu} & Y_{\Sigma_\tau}\end{array}\right)\,,
\end{equation}
and the mass matrix $M_{\Sigma}$ is now a scalar.

The second assumption we will made in the rest of this paper is to
take all the parameters real, i.e. we do not take into account the
phases of the Yukawa couplings nor the ones of the PMNS
matrix. Barring cancellations, they should not play a role in the
discovery process.

As a consequence $\epsilon$ is a $3\times3$ matrix whose elements are

\begin{equation}
\epsilon_{\alpha\beta}=\frac{v^2}{2}M_{\Sigma}^{-2}Y_{\Sigma_{\alpha}}Y_{\Sigma_{\beta}}\,,
\end{equation}

and $\epsilon'$ is now a scalar:

\begin{equation}
\epsilon'=\frac{v^2}{2}M_{\Sigma}^{-2}\left(Y_{\Sigma_e}^2+Y_{\Sigma_\mu}^2+Y_{\Sigma_\tau}^2\right).
\end{equation}

Finally, we express all the couplings in terms of the mixing
parameters,
$V_\alpha=\frac{v}{\sqrt{2}}M_{\Sigma}^{-1}Y_{\Sigma_\alpha}$, since
they are the parameters which are truly constrained by the electroweak
precision tests and the lepton flavour violating processes. Then
$\epsilon^\prime = V\cdot V^T$ while $\epsilon =V^T\wedge V$. 

By applying these simplifications and redefinitions, the couplings of
Eqs.~(\ref{gccL})-(\ref{gphiR}) in terms of $M_\Sigma$ and $V_\alpha$
are obtained; they are shown in Appendix~\ref{app:couplings}.


\section{Implementation of the model in FeynRules and validation\label{sect:implementation}}

As discussed in the previous section, the presence of an additional
fermionic triplet induces a mixing between these new heavy fermions
and the light standard model leptons. Then, not only the new couplings
must be added to the SM Lagrangian, but also SM couplings get
modified. In order to implement this model in \texttt{FeynRules}, we
start from the already implemented SM, contained in the file
\texttt{sm.fr}, and we add the new coplings and modify the existing
ones. The file containing this model is named
\texttt{typeIIIseesaw.fr}. In the following we will describe the main
features of the implemented model, before reviewing the validation
checks.

As shown before, the fermionic triplet can be expressed as a new
charged Dirac lepton $\Psi$ and a Majorana neutral lepton
$\Sigma^0$. Hence, these two new heavy particles can be viewed as a
fourth generation in the lepton sector, as suggested by the Lagrangian
and couplings written in the previous section. Therefore, a new
generation index is defined for leptons:
\begin{equation}
\texttt{IndexRange[ Index[LeptonGeneration] ] = Range[4]} ,
\end{equation}
and charged lepton and neutrino classes have to be extended to include
these new heavy particles. As for neutrinos, the whole class has to be
modified since we are now dealing with Majorana particles, while in
\texttt{sm.fr} the light neutrinos are of Dirac type~\footnote{Note
  that in the massless limit the two cases are
  equivalent.}. Consequently, the option $\texttt{SelfConjugate ->
  True,}$ is turned on. The neutrino class then reads~\footnote{The
  numbers associated to \texttt{Mass} and \texttt{Width} (for
  $\Sigma^0$) are variables.}:
\begin{eqnarray}
\texttt{F[1]} &\texttt{==}& \{ \\
&&\texttt{ClassName -> vl,}\nonumber\\
        &&\texttt{ClassMembers -> \{v1,v2,v3,tr0\},}\nonumber\\
        &&\texttt{FlavorIndex -> LeptonGeneration,}\nonumber\\
        &&\texttt{SelfConjugate -> True,}\nonumber\\
         &&\texttt{Indices -> \{Index[LeptonGeneration]\},}\nonumber\\
	&&\texttt{Mass -> \{Mv, \{Mv1, 0\}, \{Mv2, 0\}, \{Mv3, 0\}, \{Mtr0, 100.8\}\},}\nonumber\\
        &&\texttt{Width -> \{0, 0, 0, \{Wtr0, 0.1\}\}},\nonumber\\
	&&\texttt{PropagatorLabel -> \{"v", "v1", "v2", "v3","tr0"\} ,}\nonumber\\
	&&\texttt{PropagatorType -> S,}\nonumber\\
	&&\texttt{PropagatorArrow -> Forward,}\nonumber\\
        &&\texttt{PDG -> \{8000012,8000014,8000016,8000018\},}\nonumber\\
        &&\texttt{FullName -> \{"nu1", "nu2", "nu3", "Sigma0"\} \}}.\nonumber
\end{eqnarray}
Notice that, since neutrinos are Majorana particles, the kinetic term
is defined as
\begin{equation}
\texttt{I/2 vlbar.Ga[mu].del[vl, mu]}\,.
\end{equation}
Analogously, the charged leptons class now reads:
\begin{eqnarray}
\texttt{F[2]} &\texttt{==}& \{ \\
        &&\texttt{ClassName -> l,}\nonumber\\
        &&\texttt{ClassMembers -> \{e, m, tt,trm\},}\nonumber\\
        &&\texttt{FlavorIndex -> LeptonGeneration,}\nonumber\\
        &&\texttt{SelfConjugate -> False,}\nonumber\\
        &&\texttt{Indices -> \{Index[LeptonGeneration]\},}\nonumber\\
	&&\texttt{Mass -> \{Ml, \{Me, 5.11 * 10(-4)\}, \{MM, 0.10566\}, \{MTA, 1.777\}, \{Mtrch, 101\}\},}\nonumber\\
        &&\texttt{Width -> \{0, 0, \{Wtau, 0.1\}, \{Wtrch, 0.1\}\},}\nonumber\\
	&&\texttt{QuantumNumbers -> \{Q -> -1\},}\nonumber\\
	&&\texttt{PropagatorLabel -> \{"l", "e", "m", "tt", "tr-"\},}\nonumber\\
	&&\texttt{PropagatorType -> Straight,}\nonumber\\
        &&\texttt{ParticleName ->\{"e-", "m-", "tt-", "tr-"\},}\nonumber\\
        &&\texttt{AntiParticleName -> \{"e+", "m+", "tt+", "tr+"\},}\nonumber\\
	&&\texttt{PropagatorArrow -> Forward,}\nonumber\\
        &&\texttt{PDG -> \{11, 13, 15,8000020\},}\nonumber\\
       &&\texttt{FullName -> \{"Electron", "Muon", "Tau", "Sigma-"\} \}}.\nonumber
\end{eqnarray}

Notice that the usual PDG codes for light neutrinos (12, 14, 16) have
been replaced by new codes (8000012, 8000014, 8000016), since in our
model light neutrinos are no longer Dirac particles but Majorana
ones. Moreover new codes have been provided for the neutral component
(8000018) and the charged component (8000020) of the triplet. These
codes are currently not officially used for other particles species
and any change should be done very carefully not to interfere with
existing assignments (see Particle Data Group numbering Scheme
\cite{Nakamura:2010zzi}).

Having (re)defined the lepton fields, the interactions can be
implemented in the Lagrangian. Since the light leptons couplings to
the gauge bosons and Higgs fields are different from the SM case, they
have been erased and replaced by the ones defined in the previous
sections. The matrices
$g^{CC}_{L/R},\,g^{NC}_{L/R},\,g^{H_\nu}_{L/R},\,g^{H_l}_{L/R}$ and
$g^{\phi^-}_{L/R}$ defining the couplings have been introduced as
internal parameters in order to write the Lagrangian in a clear
way. The external parameters, or inputs, are listed in
Table~\ref{tab:inputs}. In this table some values for the parameters
of the model implemented in \texttt{typeIIIseesaw.fr} are given, but
these are variables that can be modified according to the details of
the considered model.

Following the features of the SM implementation, our model presents
the characteristic of allowing a differentiation between the kinematic
mass (or pole mass) of the triplet and the masses entering into the
couplings definition (equivalent of Yukawa masses). The former are
defined under the block MASS while the latter are defined under the
block NEWMASSES. In particular, for the charged fermion masses, we
have made the same assignments as in \texttt{sm.fr}: the Yukawa masses
for $e$, $\mu$, $u$, $d$, $s$ are zero while their pole masses, which
are used for example by \texttt{PYTHIA}, are non-zero. This implies
that any coupling defined in terms of the Yukawa masses will be zero
in our model. We have checked that turning on this Yukawa masses would
amount to a negligible correction.


\subsection{Validation}

In this section we discuss the checks we have performed in order to
validate the model we have implemented by comparing some numerical
results on branching ratios and cross sections obtained with
\texttt{typeIIIseesaw.fr} and \texttt{sm.fr}. Moreover, when possible,
we will compare the numerical results with some analitic
expressions. In Table~\ref{tab:inputs} the list of the parameters used
for the comparison is given.

\begin{table}[!h]
\begin{center}
\begin{tabular}{|r|c|c|c|}
\hline
Parameter & Symbol & Value in \texttt{sm.fr} & Value in \texttt{typeIIIseesaw.fr} \\
\hline
\hline
Inverse of the electromagnetic coupling & $\alpha_{EW}^{-1}(M_Z)$ & 127.9 & 127.9\\
Strong coupling  & $\alpha_s(M_Z)$ & 0.118 & 0.118\\
Fermi Constant  & $G_F$ & 1.16639e-5 GeV$^{-2}$ & 1.16639e-5 GeV$^{-2}$\\
\hline
 Z pole mass & $M_Z$ & 91.188 GeV & 91.188 GeV\\
 c quark mass & $m_c$ & 1.42 GeV & 1.42 GeV\\
 b quark mass & $m_b$ & 4.7 GeV & 4.7 GeV\\
 t quark mass & $m_t$ & 174.3 GeV & 174.3 GeV\\
 $\tau$ lepton mass & $m_{\tau}$ & 1.777 GeV & 1.777 GeV\\
 Higgs mass & $M_H$ & 120 GeV & 120 GeV\\
 Cabibbo angle & $\theta_c$ & 0.227736 & 0.227736\\
 \hline
 Electron mass & $m_e$ & 0 & 0\\
 Muon mass & $m_{\mu}$ & 0 & 0\\
Charged heavy fermion  mass & $M_{\Sigma}$ & - & 101 GeV\\
 Neutral heavy fermion mass & $M_{\Sigma^0}$ & - & 100.8 GeV\\
 Light neutrino mass  & $m_1$ & 0 & 0\\
 & $m_2$ & 0 & 0\\
 & $m_3$ & 0 & 0\\
 PMNS mixing angles $\theta_{12}$ & $\theta_{12}$ & - & 0.6\\
 $\theta_{23}$ & $\theta_{23}$ & - & 0.75 \\
 $\theta_{13}$ & $\theta_{13}$ & - & 0.1 \\
 Heavy-light fermion mixing $V_e$ & $V_e$ & - & 0\\
 $V_{\mu}$ & $V_{\mu}$ & - & 0.063 \\
 $V_{\tau}$ & $V_{\tau}$ & - & 0 \\ 
\hline 
\end{tabular}
\end{center}
\caption{Input parameters for \texttt{sm.fr} and \texttt{typeIIIseesaw.fr}.}
\label{tab:inputs}
\end{table}

We start by comparing some branching ratios that should not be
affected (or very slightly) by the presence of the triplet between the
\texttt{FeynRules} unitary-gauge implementations in
\texttt{MadGraph/MadEvent} of the Type III seesaw
(\texttt{typeIIIseesaw\_MG}) and the SM (\texttt{sm\_FR}). These
branching ratios have been calculated with the program
\texttt{BRIDGE}~\cite{Meade:2007js}~\footnote{Some care has to be
  taken when calculating branching ratios with \texttt{BRIDGE} with
  Majorana particles. Here the branching ratios for Z going into
  Majorana particle has been fixed ``by hand''.} and are gathered in
Table~\ref{tab:validation-Br} in Appendix~\ref{app:validation}. They
agree within $1.5\%$ which roughly corresponds to the intrinsic error
of this program; the deviation induced by the presence of the triplet
is indeed much smaller ($\sim 0.3\%$).

\begin{figure}[!t]
\begin{center}
\epsfig{figure=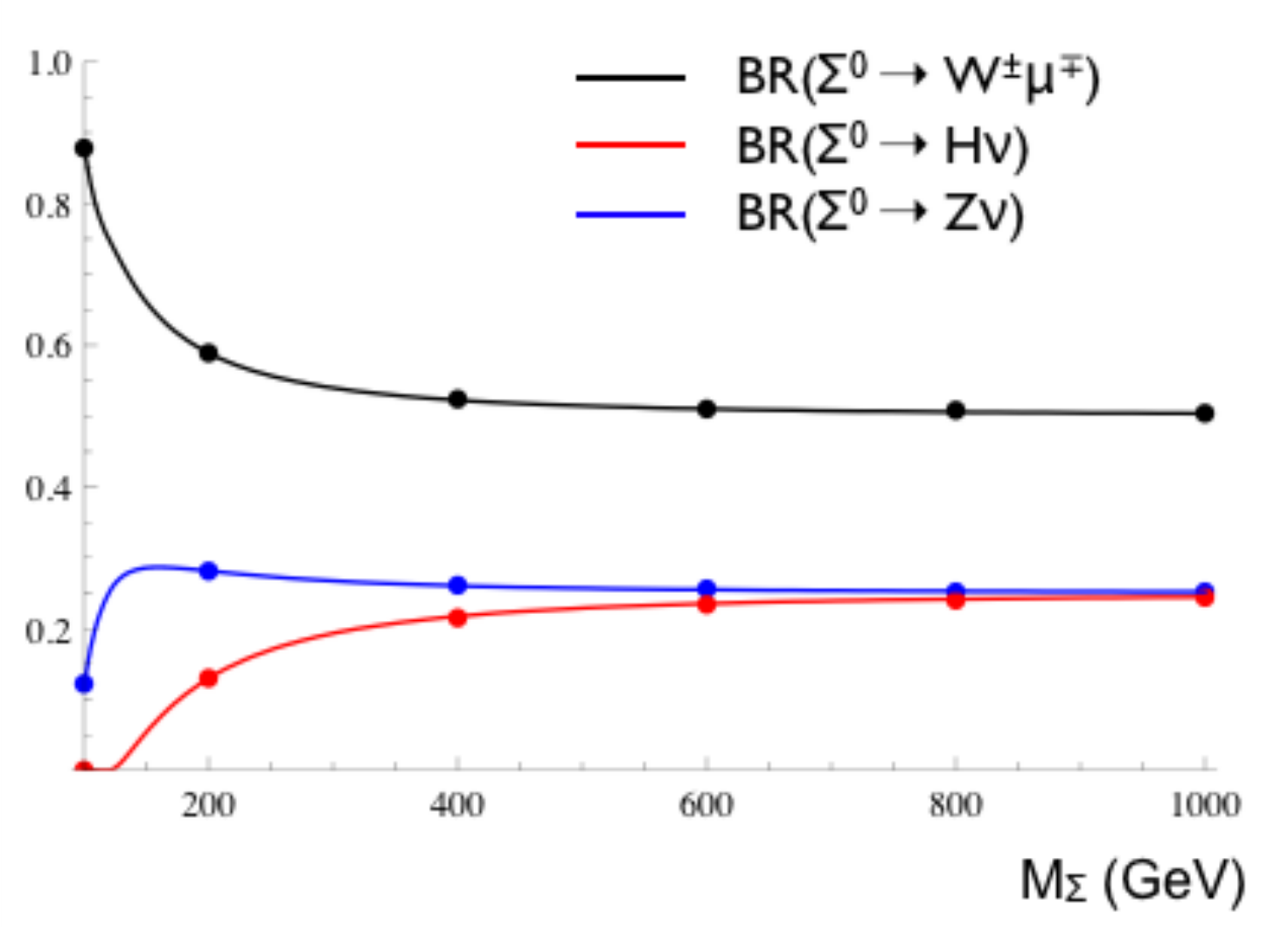,width=8cm}
\hspace*{0.5cm}
\epsfig{figure=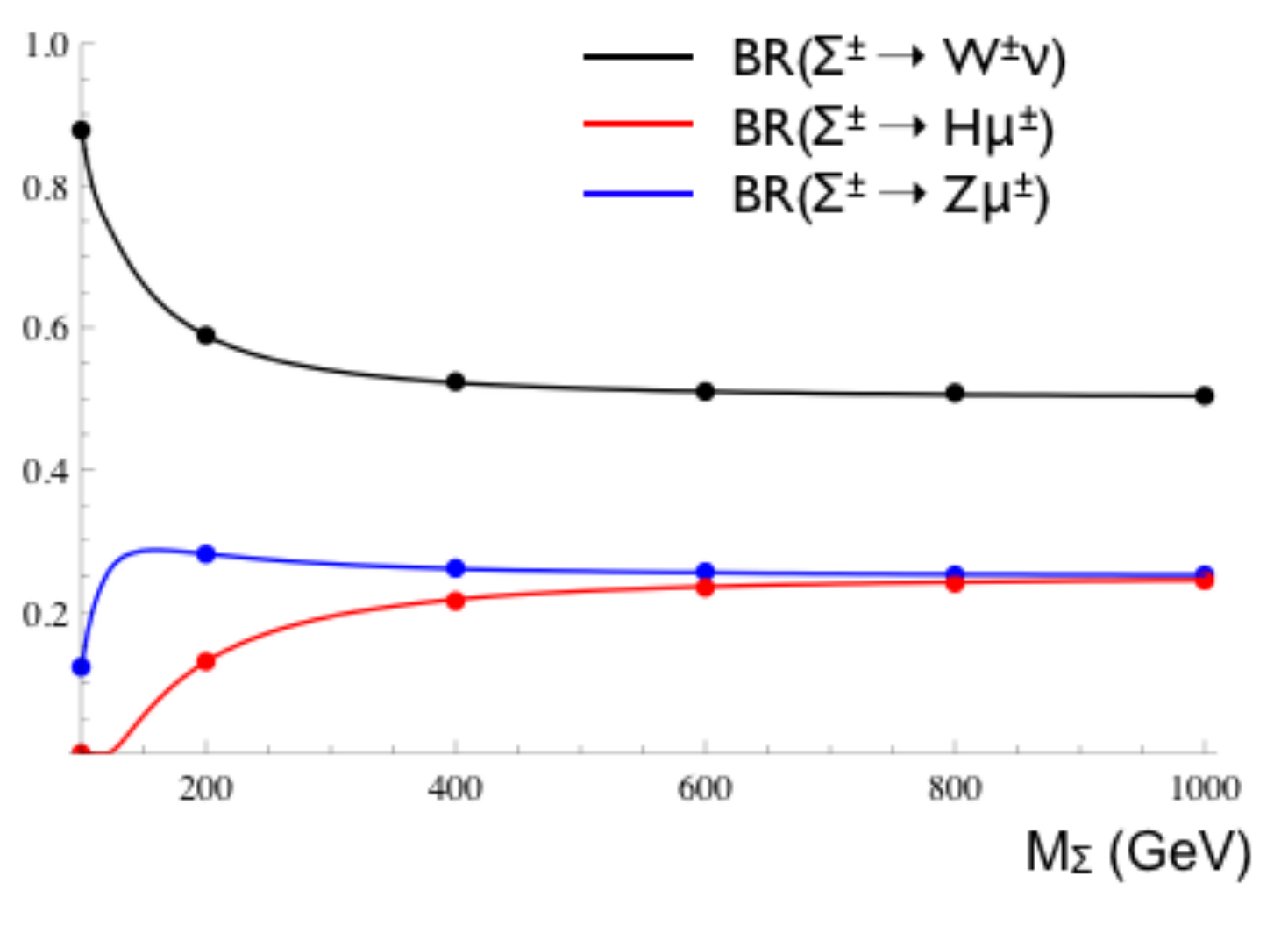,width=8cm}
\caption{Branching ratios of the neutral component (left) and charged
  component (right) of the fermionic triplet in the case
  $V_e=V_\tau=0\, ,\ V_\mu = 0.063$. The dots correspond to
  numerically evaluated values while the lines correspond to the
  theoretical predictions. Notice that, as expected from
  Eqs.~(\ref{eq:Gammatr0W})-(\ref{eq:Gammatr+h}) in the case of one
  non-zero mixing angle, the result is the same for charged and
  neutral triplet decay.}
\label{fig:BR_Vmu} 
\end{center}
\end{figure}

\begin{figure}[!h]
\begin{center}
\epsfig{figure=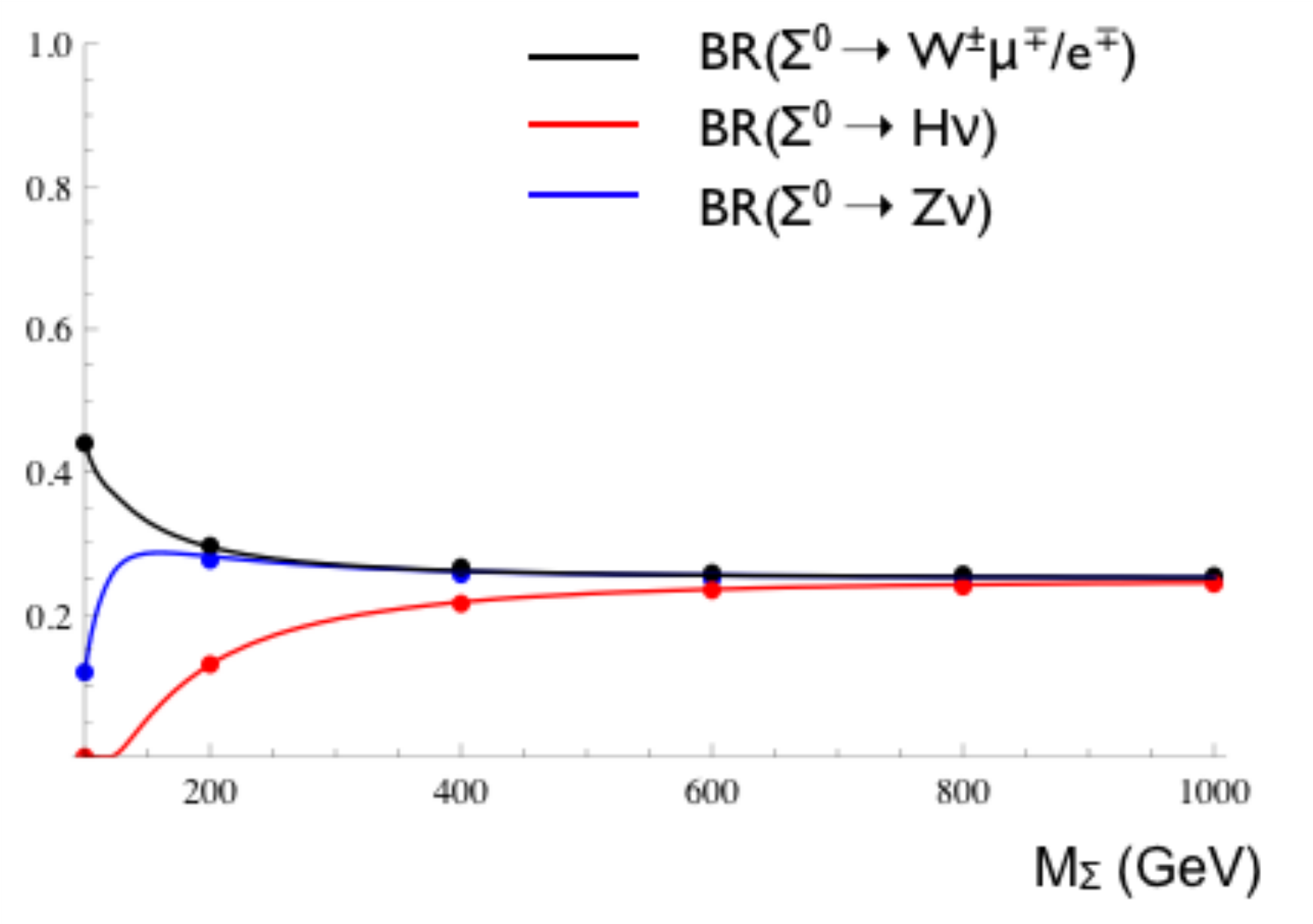,width=8cm}
\hspace*{0.5cm}
\epsfig{figure=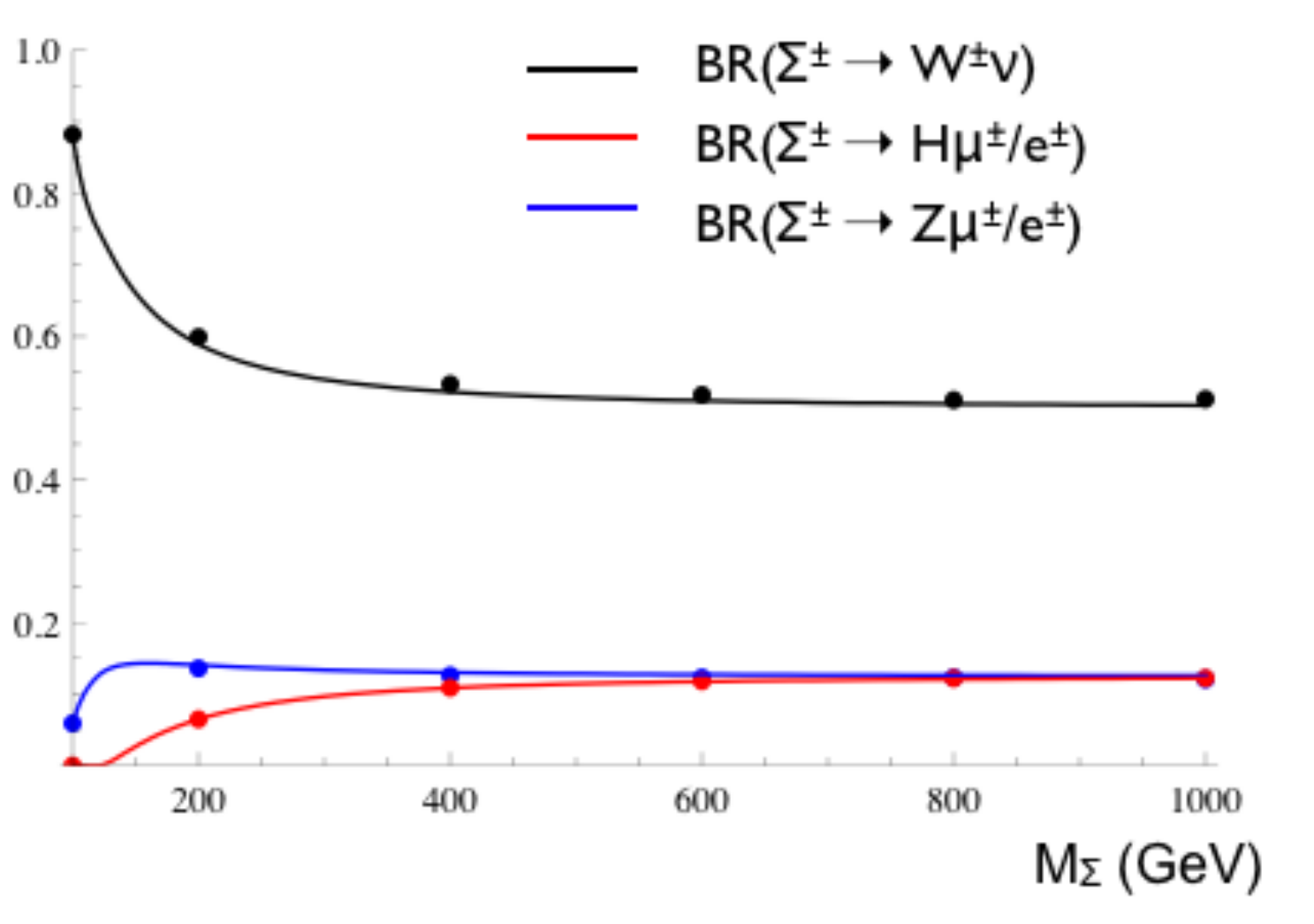,width=8cm}
\caption{Branching ratios of the neutral component (left) and charged
  component (right) of the fermionic triplet in the case $V_\tau=0\,
  ,\ V_e=V_\mu = 4.1\cdot 10^{-4}$. The dots correspond to numerically
  evaluated values while the lines correspond to the theoretical
  predictions while the lines correspond to the theoretical
  predictions. When both channel with $e$ and $\mu$ are open, only one
  is displayed, since, for this particular choice of the mixing
  angles, they are overlapped.}
\label{fig:BR_VeVmu} 
\end{center}
\end{figure}

Additionally, these branching ratios can be confronted with the
analitic expressions that can be derived from the following decay
width~\cite{Franceschini:2008pz}:
\begin{eqnarray}
\label{eq:Gammatr0W}
\Gamma(\Sigma^0 \to l_\alpha^- W^+) & = &  \Gamma(\Sigma^0 \to l_\alpha^+ W^-) =
\frac{g^2}{64 \pi} |V_\alpha|^2
\frac{M_\Sigma^3}{M_W^2} \left( 1- \frac{M_W^2}{M_\Sigma^2} \right)^2 
\left( 1 + 2 \frac{M_W^2}{M_\Sigma^2} \right) , 
\\[0.1cm]
\label{eq:Gammatr0Z}
\sum_l\Gamma(\Sigma^0 \to \nu_l Z) & = &  \frac{g^2}{64 \pi c_W^2} 
\sum_\alpha |V_\alpha|^2
\frac{M_\Sigma^3}{M_Z^2} \left( 1- \frac{M_Z^2}{M_\Sigma^2} \right)^2 
\left( 1 + 2 \frac{M_Z^2}{M_\Sigma^2}\right),
\\[0.2cm]
\label{eq:Gammatr0h}
\sum_l\Gamma(\Sigma^0 \to \nu_l H) & = &  \frac{g^2}{64 \pi} 
\sum_\alpha |V_\alpha|^2
\frac{M_\Sigma^3}{M_W^2} \left( 1- \frac{M_H^2}{M_\Sigma^2} \right)^2 ,
\\[0.2cm]
\label{eq:Gammatr+W}
\sum_l\Gamma(\Sigma^+ \to \nu_l W^+) & = & 
\frac{g^2}{32 \pi} \sum_\alpha |V_\alpha|^2
\frac{M_\Sigma^3}{M_W^2} \left( 1- \frac{M_W^2}{M_\Sigma^2} \right)^2 
\left( 1 + 2 \frac{M_W^2}{M_\Sigma^2} \right) ,
\\[0.2cm]
\label{eq:Gammatr+Z}
\Gamma(\Sigma^+ \to l_\alpha^+ Z) & = & 
\frac{g^2}{64 \pi c_W^2} |V_\alpha|^2
\frac{M_\Sigma^3}{M_Z^2} \left( 1- \frac{M_Z^2}{M_\Sigma^2} \right)^2 
\left( 1 + 2 \frac{M_Z^2}{M_\Sigma^2}\right),
\\[0.2cm]
\label{eq:Gammatr+h}
\Gamma(\Sigma^+ \to l_\alpha^+ H) & = & \frac{g^2}{64 \pi} |V_\alpha|^2
\frac{M_\Sigma^3}{M_W^2} \left( 1- \frac{M_H^2}{M_\Sigma^2} \right)^2 .
\end{eqnarray}
Fig.~\ref{fig:BR_Vmu} shows the branching ratios of the charged and
neutral component of the fermionic triplet in the case $V_e=V_\tau=0\,
,\ V_\mu = 0.063$, while Fig.~\ref{fig:BR_VeVmu} shows the branching
ratios in the case $V_\tau=0\, ,\ V_e=V_\mu = 4.1 \cdot 10^{-4}$. In
both figures, the dots represent the values calculated by
\texttt{BRIDGE} while the lines correspond to the theoretical
predictions. A great agreement is evident.

Notice that, in case of small mixing angles, the three-body decays of
$\Sigma^+$ into $\Sigma^0\ e^+(\mu^+)\ \nu$ and especially into
$\Sigma^0\ \pi^+$ could become relevant~\cite{Franceschini:2008pz} and
should be taken into account when computing branching ratios. We have
checked that, for mixing angles of the order of $10^{-6}$,
$\textrm{Br}(\Sigma^+ \to\Sigma^0\ \pi^+)\sim 10^{-3}$, i.e. 2 orders
of magnitude smaller than other dominant decays.



As a second step of the validation procedure, we have computed the
cross sections of a selection of $2\rightarrow 2$ processes that
should not be influenced by the presence of triplets using
\texttt{MadGraph/MadEvent} and we have compared the results obtained
with \texttt{typeIIIseesaw\_MG} and \texttt{sm\_FR}. Results are
gathered in Table~\ref{tab:validation-2to2} in
Appendix~\ref{app:validation}: an agreement at the level of $1\%$ is
found.

Finally, we have checked that the production of a pair of triplets at
the LHC with a center-of-mass energy of 14~TeV obtained with
\texttt{MadGraph/MadEvent} matches the previous results in the
literature~\cite{delAguila:2008cj,Franceschini:2008pz}, see
Table~\ref{tab:prod-14TeV} in Appendix~\ref{app:validation}.


\section{The minimal type III seesaw model at the LHC at 7~TeV}


\subsection{Bounds on the mixing angles}

In Refs.~\cite{Abada:2008ea,Abada:2007ux,delAguila:2008pw} the bounds on the parameters
of the type III seesaw model have been derived. The bounds apply to
the following combination of parameters:
\begin{equation}
  \label{eq:boundepsilon}
  \frac{v^2}{2} \left| Y^\dagger M^{-2}Y \right|_{\alpha\beta}= \left|
    V_\alpha V_\beta \right|.
\end{equation}
We have then the following constraints:
\begin{eqnarray}
 |V_e|&<& 5.5\cdot 10^{-2} 
\label{e}\\
 |V_\mu| &<& 6.3\cdot 10^{-2} 
\label{mu}\\
 |V_\tau| &<&  6.3\cdot 10^{-2} 
\label{tau}\\
 |V_e V_\mu| &<& 1.7\cdot 10^{-7} 
\label{emu}\\
 |V_e V_\tau| &<& 4.2\cdot 10^{-4}
\label{etau}\\
 |V_\mu V_\tau|&<& 4.9\cdot 10^{-4}.
\label{mutau}
\end{eqnarray}

Notice that if only $V_e$ or $V_\mu$ is present the stronger constrain
of Eq.~(\ref{emu}) does not apply and $\mathcal{O}(10^{-2})$ mixings
are allowed. On the other side, if both are different from zero, then
either one of the two is much smaller than the other, effectively
reducing this case to the one with only one non-zero $V_\alpha$, or
they are both $\mathcal{O}(10^{-3})$, in order to satisfy the strong
bound of Eq.~(\ref{emu}).  However, as we will discuss later, since
the production of the triplet happens via gauge interactions, reducing
the mixing angle will not reduce the total cross section, so that
these bounds have to be taken into account, but the mixing angles are
not as crucial as in the type I seesaw.

In this paper we are going to focus on a specific case, in order to
illustrate how our model works and to show that even with the LHC
running at 7~TeV there is the possibility of testing the low scale
type III seesaw. We are going to give the cross section of the
relevant channels for the case $V_e=V_\tau=0 \ , \ V_\mu =
0.063$. This case corresponds to the maximum allowed mixing angles. If
the mixing is so large, then some cancellation or an extended seesaw
mechanism like the inverse seesaw must be invoked in order to obtain
the correct value for neutrino masses. However, all the discussion we
perform in this section applies also in the case of small mixing. In
the next sections we are going to discuss the triplet production and
decays, give the cross sections which are relevant for discovery and
discuss the main backgrounds which affect the measurement and the main
cuts that could be implemented in order to reduce it. A more detailed
study is beyond the scope of this work.


\subsection{Triplet production and decay}

At the LHC triplets are mainly produced in pair. In
Table~\ref{tab:prod-7TeV} production cross sections for different mass
values are collected, with the acceptance cuts listed in
Table~\ref{tab:acceptance-cuts}.  Since the triplets are produced via
gauge interactions, the production cross sections do not depend on the
mixing parameters.  After production, the triplets decay inside the
detector according to the expressions displayed in
Eqs.~(\ref{eq:Gammatr0W})-(\ref{eq:Gammatr+h}). While the decay width
depends strongly on the value of the mixing angles $V_\alpha$, the
branching ratios dependence is very mild. Since we are always in the
narrow width regime, the total cross section is driven only by the
mass of the triplet (for the production) and its branching ratios (for
the decays).  Therefore, a non-discovery at the LHC will permit to
constrain the mass of the triplet, after some assumption on the
branching ratios have been done.

Once the triplets have decayed into leptons and gauge bosons, the
latter will then decay into charged leptons, quarks, which will show
up as jets (and leptons, when heavy quarks decay semileptonically),
and neutrinos, which will manifest themselves as missing energy. Final
states can be classified according to the number of charged
leptons. The type III seesaw can give rise to final states with up to
6 leptons. However, it has been shown that the cross sections for 6-,
5- and 4-leptons final states is to low for being useful for
discovery, already at 14~GeV~\cite{delAguila:2008cj}; therefore, we
will not consider them here~\footnote{However, since the probability of
  missing a lepton is relatively high for multilepton channels, when
  generating events to study the possibility of having a signal in the
  3- and 2-leptons channels, events with 4 leptons should be generated
  too. The inclusive 4-leptons final state cross section varies
  between 10-20~fb for triplet masses in the range 100-140~GeV.}. On
the other hand, the most promising channels are the 3-leptons and the
dileptons, i.e. with 2 leptons of the same sign. In the following
sections we are going to discuss these channels and the main
backgrounds which affect them.

\begin{table}[!h]
\begin{center}
\begin{tabular}{|l||l|l|l|}
\hline
$M_\Sigma$ & $\sigma(pp\to \Sigma^+ \Sigma^0)(fb)$ 
& $\sigma(pp\to \Sigma^+ \Sigma^-)(fb)$ & $\sigma(pp\to \Sigma^- \Sigma^0)(fb)$ \\
\hline
\hline
100 & 4.329e+3&3.339e+3&2.325e+3 \\
\hline
120 & 2.157e+3 & 1.629e+3 & 1.106e+3  \\
\hline
140 & 1.200e+3 & 8.882e+2 & 5.894e+2  \\
\hline
160 & 7.215e+2 & 5.229e+3 & 3.387e+2 \\
\hline
180 & 4.555e+2 & 3.249e+2 & 2.059e+2 \\
\hline
200 & 3.006e+2& 2.109e+2 & 1.311e+2 \\
\hline
300 & 5.488e+1&3.580e+1&2.027e+1 \\
\hline
400 & 1.434e+1&8.777 & 4.632 \\
\hline
600 & 1.527& 8.576e-1 &4.118e-1 \\
\hline
800 & 2.097e-1 & 1.132e-1 & 5.139e-2 \\
\hline
1000 & 3.133e-2 & 1.774e-2 & 7.401e-2 \\
\hline 
\end{tabular}
\end{center}
\caption{Production cross sections at 7~TeV. }
\label{tab:prod-7TeV}
\end{table}

\begin{table}[!h]
\begin{center}
\begin{tabular}{|c|c|c|}
\hline
\multicolumn{3}{|c|}{Acceptance Cuts} \\
\hline
\hline
$p_{T_j}>20\,\rm{GeV}$ & $\eta_j<5$ & $\Delta R_{jj}>0.001$\\
\hline
$p_{T_\ell}>10\,\rm{GeV}$ & $\eta_\ell<2.5$ & $\Delta R_{\ell\ell}>0$\\
\hline 
\end{tabular}
\end{center}
\caption{Acceptance cuts used for production simulations at 7~TeV and 14~TeV. }
\label{tab:acceptance-cuts}
\end{table}


\subsection{The most relevant final states}

Tables~\ref{tab:dileptons} and \ref{tab:trileptons++-} in
Appendix~\ref{app:cross} display the cross sections for the
intermediate and final states with 2 and 3 leptons at different mass
energies.~\footnote{We give numbers for the case of mixing with muons
  exclusively, however similar results apply when the final states
  contains electrons as well. On the other hand, they do not apply
  completely to taus. Indeed, taus are not detected as such, because
  of their fast decay. Moreover, in a detector like CMS, leptons
  coming from taus decay are not distinguished from prompt leptons and
  therefore identified taus are only hadronic taus.} While the
intermediate ones are calculated with \texttt{MadGraph}, the final
ones are obtained by multiplication with the corresponding branching
ratios. From a quick look to these tables one can see that even with
LHC running at 7~TeV, with the few fb$^{-1}$ of luminosity which are
expected to be reached by the end of 2011, several events are
expected, for low triplet mass. In the 3-leptons table, in the total
cross section we have isolated the channels with leptons not-coming
from Z decay. Indeed, when the cut on the invariant mass of the
leptons will be applied in order to reduce the background events
coming from Z decay (see later), these events will mostly
disappear. Then the numbers we quote in blue in
Table~\ref{tab:trileptons++-} can be considered the effective cross
section after the application of this cut.

By looking at these table we see that there are 4 possible final
states with 2 and 3 leptons:
\begin{itemize}
\item[{\bf A)}] 3 leptons + missing transverse energy (MET);
\item[{\bf B)}] 3 leptons + 2 jets + MET;
\item[{\bf C)}] 2 same-sign leptons + 4 jets;
\item[{\bf D)}] 2 same-sign leptons + 2 jets + MET.
\end{itemize}
In what follows we are going to discuss the main features of all of
them. We have simulated $pp\rightarrow \Sigma^+ \Sigma^0 \rightarrow
\mu^+\mu^+\mu^- + \nu s (+jets)$ with \texttt{MadGraph/MadEvents},
hadronization being obtained with the help of
\texttt{PYTHIA}~\cite{PYTHIA}. The CMS detector has been simulated via
the PGS software~\cite{PGS}.

\begin{figure}[!t]
\begin{center}
\epsfig{figure=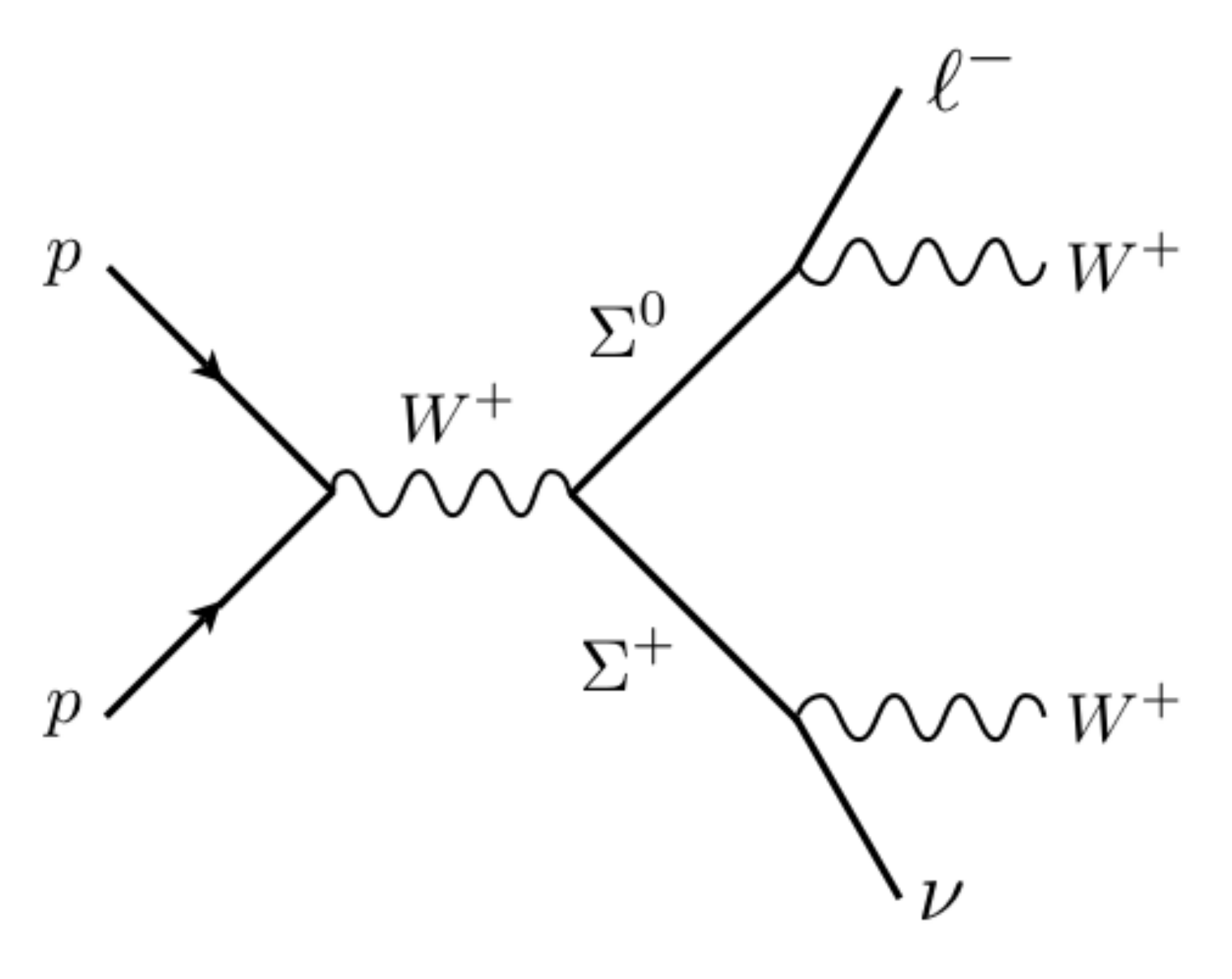,width=6cm}
\caption{Dominant process for the discovery channel for the fermionic triplet at the LHC.}
\label{fig:Process}
\end{center}
\end{figure}

\begin{figure}[h]
\begin{center}
\epsfig{figure=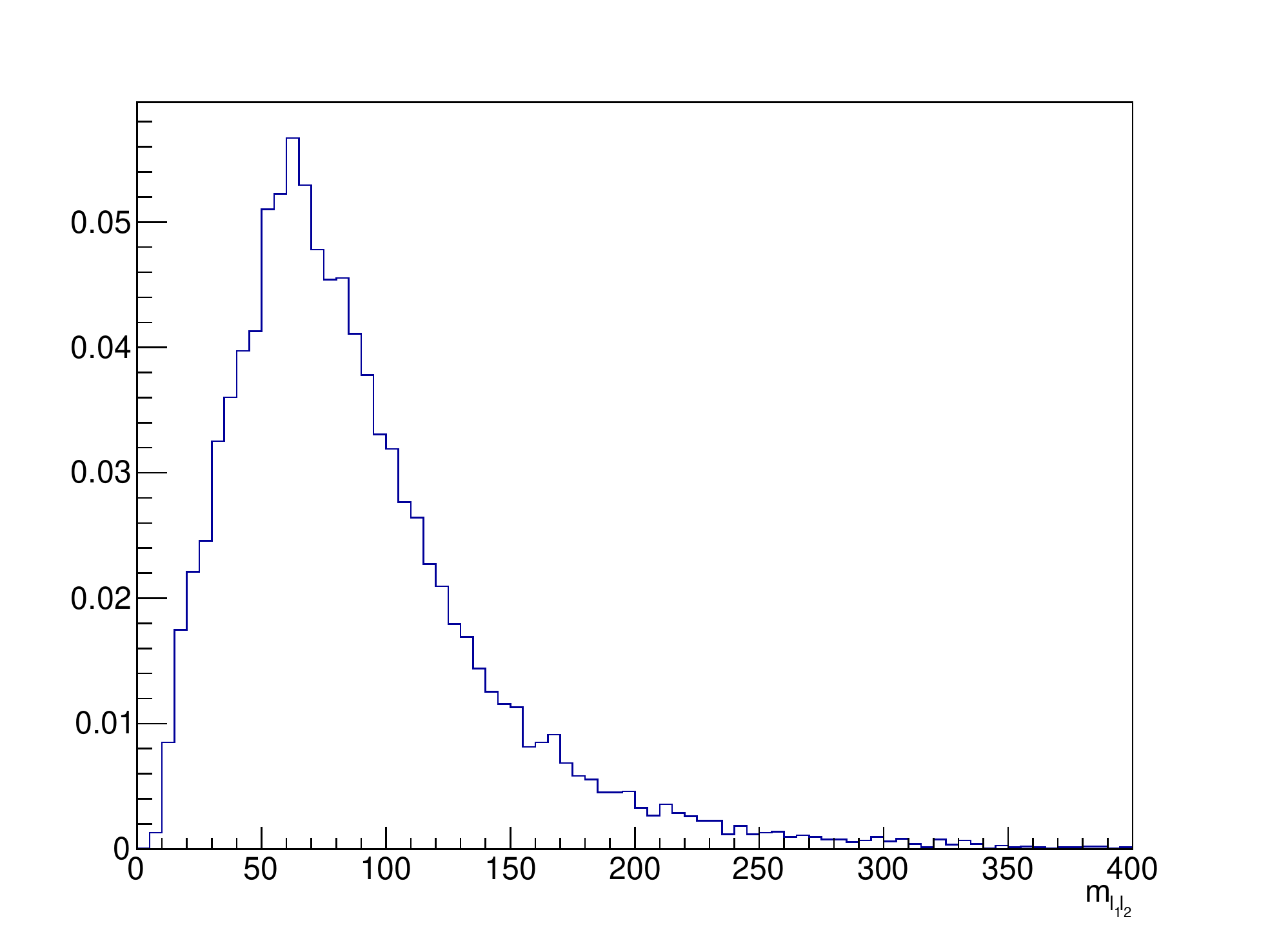,width=8cm}
\caption{Invariant mass of the two $\mu^+$ for a luminosity of
  $30\rm{fb}^{-1}$ and $M_\Sigma=100~\rm{GeV}$. Pre-selection cuts
  selected only the events with 3 charged leptons among which 2
  positive muons.}
\label{fig:invmass} 
\end{center}
\end{figure}

\begin{description}
\item{\bf 3 leptons + MET.} This is probably the best discovery
  channel: indeed the background is more easily reduced due to the
  absence of jets in the final state. The dominant process generating
  it is depicted in Fig.~\ref{fig:Process}. In an ideal detector where
  jets are not misidentified with leptons, the only background sources
  would be $WW$, $WWW$, $WZ$ and $ZZ$ when a lepton is missed. In
  practice jets should be added to these background; however, as it is
  discussed later, all these background should be under control.

 In this channel, the invariant mass $m_{\mu^+ \mu^+}$ of the two
 same-sign muons presents a long tail in the high energy region that is
 characteristic of the presence of new physics, see
 Fig.~\ref{fig:invmass}, and can be exploited to reduce the
 background. Moreover, this is typical of this kind of seesaw,
 permitting thus to distinguish among type I, II and
 III~\cite{delAguila:2008cj}.

\begin{figure}[!h]
\begin{center}
\epsfig{figure=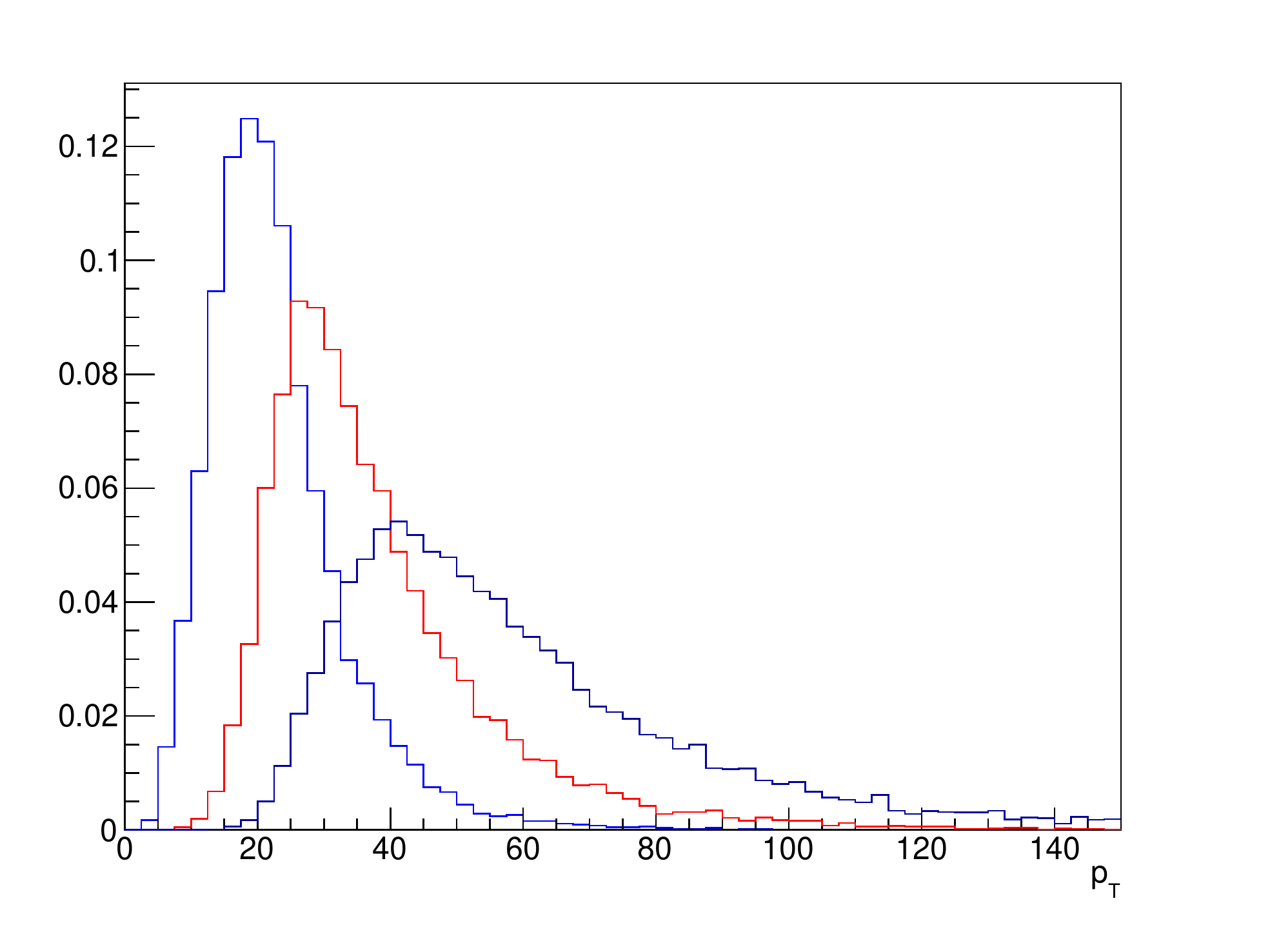,width=9cm}
\caption{\label{fig:pt} $p_T$ distribution of the different leptons
  for $M_\Sigma=100~\rm{GeV}$. The black, red and blue curves
  represent the lepton with the highest, intermediate and smallest
  $p_T$ respectively. Pre-selection cuts selected only the events with
  3 charged leptons among which 2 positive muons.}
\label{fig:pt_3lMET}
\end{center}
\end{figure}
 
%
\item{\bf 3 leptons + 2 jets + MET.} This channel is probably the best
  one in order to reconstruct the mass of the triplet. Moreover it can
  be used also to discriminate between type II and type III
  seesaw~\cite{delAguila:2008cj}. It also appears in the type I seesaw
  with a gauged $U(1)_{B-L}$~\cite{basso}. In this case the reduction
  of the background can be more complicated, due to the impossibility
  of applying a jet veto. Essentially all the sources listed in the
  next section constitute a background for this channel. A precise
  estimation of the sensitivity to this new physics would require the
  complete simulation of the background and a detailed analysis, which
  is beyond the scope of this work. However, we will show later that
  the possibility of reducing the background to ``reasonable'' levels
  is realistic.

Once the triplet has been observed, its mass needs to be measured. To
this aim, this channel, emerging from the process $pp\rightarrow
(\Sigma^{\pm}\rightarrow \ell^{\pm} Z/H)(\Sigma^0 \rightarrow
\ell^{\pm}W^{\mp})$ with $Z/H$ decaying into jets, is the best
one. Indeed the momentum of the $Z/H$ boson is reconstructed from the
jets momenta, while its combination with the momentum of one of the
two same-sign leptons gives the mass of the charged triplet. Since
there are two possibilities for this combination, the chosen one will
be that giving closest invariant mass for the recontructed charged and
neutral triplets, where the latter is given by the combination of the
momenta of the two remaining leptons $plus$ MET~\footnote{The neutrino
  longitudinal momento should be added as
  well~\cite{delAguila:2008cj}.}.

The reconstructed mass of the charged and neutral triplet are shown in
Fig.~\ref{fig:massrec} where no cuts has been applied. Note that a
selection cut on the invariant mass $m_{jj}$ of the jets
\begin{equation}
|m_{jj}-M_{Z/H}|< 10\,\rm{GeV}
\end{equation}
will improve the mass reconstruction.
%
%
\begin{figure}[!h]
\begin{center}
\epsfig{figure=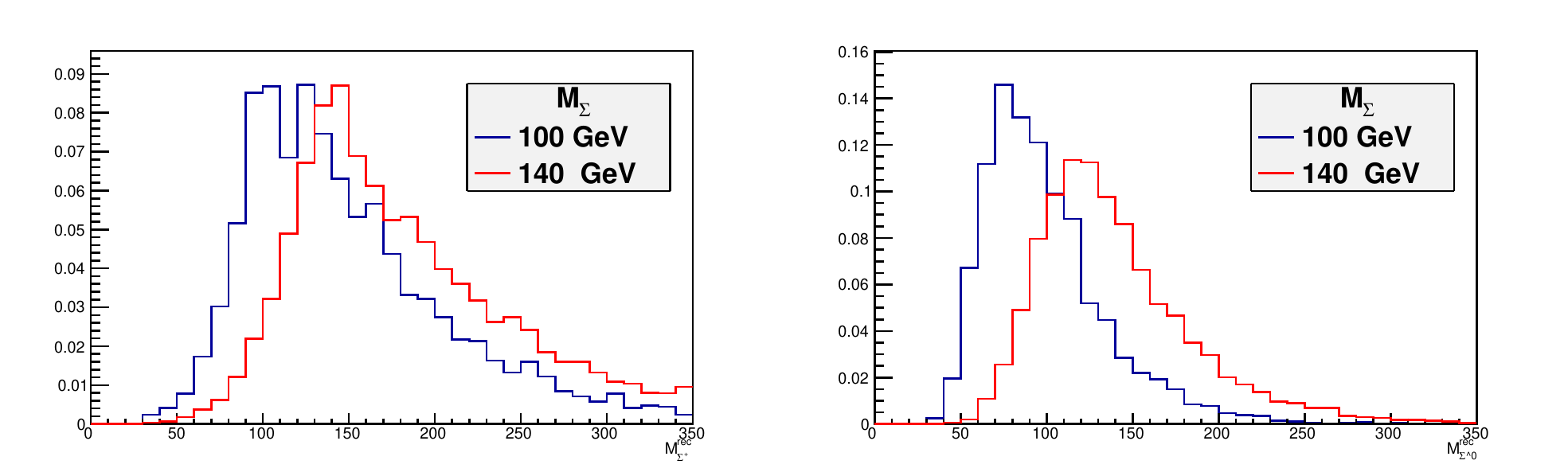,width=18cm}
\caption{\label{fig:massrec} Reconstructed mass of the charged triplet
  (left) and neutral triplet (right), for a luminosity of
  $30\rm{fb}^{-1}$, in the case $M_\Sigma=100~\rm{GeV}$ (black curve)
  and $M_\Sigma=140~\rm{GeV}$ (red curve). Pre-selection cuts selected
  only the events with 3 charged leptons and at least 2 jets.}
\end{center}
\end{figure}
%
%
Even if the background is added, a clear peak in the reconstructed
mass will still be visible, which should also permit to distinguish
from type II seesaw~\cite{delAguila:2008cj}.
\item{\bf 2 same-sign leptons + jets (+MET)} As it is clear from
  Table~\ref{tab:dileptons}, the cross section for these final states
  are quite large, even larger than the ones for 3 leptons final
  states. However here jets are always present, which can render a bit
  more difficult the background reduction. The backgrounds are
  essentially the same as in the previous channel and indeed it has
  been shown~\cite{delAguila:2008cj} that the discovery and the
  discriminatory potentials of the 2- and 3-leptons final states are
  similar too. A realistic study, especially a study on real data,
  should consider this channel as well.
\end{description}


\subsection{Background}

The main background sources for the channels discussed above are :
$t\overline{t}$, $t\overline{t}W$, $WW$, $WZ$, $ZZ$, $Zt\overline{t}$,
$Zb\overline{b}$ and 3 gauge bosons. The same background plus
additional jets should be considered as well, both if looking at final
states with jets or no: some jets can be indeed misidentified as
leptons.  In the following we will give a brief description of each
background and of the cuts that can be implemented in order to reduce
it. Whenever the cross section for the different background under
study has not been measured, we have used \texttt{MadGraph/MadEvent}
to obtain the cross-sections for LHC running at 7~TeV and compared our
results with previous results obtained by the CMS
collaboration~\cite{Chatrchyan:2011em} whenever possible. All
backgrounds have been simulated with 0 and 1 additional jets.
 \begin{description}
 \item{$\mathbf{ t\overline{t}.}$} The production of a pair of top
   quarks decaying into $bW$, one of the $b$ giving a lepton and the W
   decaying leptonically, is a source of background with a large cross
   section. At 7~TeV the production of a top quarks pair has been
   measuread by CMS~\cite{ttbar1} and ATLAS~\cite{ttbar2} to be
   $\sigma_{t\overline{t}}=(173^{+39}_{-32})\,\rm{pb}$ and
   $(171\pm20\pm14^{+8}_{-6})\,\rm{pb}$, with an integrated luminosity
   of 36 and 35 pb$^{-1}$, respectively. Combining the branching ratio
   $BR(W\rightarrow l\nu)=30\%$ with the $10\%$ of branching ratio for
   the semileptonic decay of the $b$, the final cross section for such
   background should be around $0.15-1.5\,\rm{pb}$ depending on how
   many different lepton flavors one expect in the final state. In the
   case where the signal final state does not contain jets (at the
   parton level), a cut on the number of jets will reduce this
   background to negligible levels. $b$-tagging could be applied in
   order to reduce it when channels with jets are considered.
 \item{$\mathbf{t\overline{t}W.}$} Here the two tops decay into a $W$
   plus jets. The third $W$ ensures the presence of three leptons in the
   final state. The presence of jets makes this background negligible
   when looking to three leptons + MET without jets. On
   the other hand, when channels with jets are considered, this
   background should be carefully studied. We found
   $\sigma_{ttWj}\sim 230~\rm{fb}$. The
   production cross section for $t\bar{t}W$ should then be larger, but
   considering the appropriate branching fractions, the final cross
   sections should be of few fb, depending on the number of jets.
 \item{$\mathbf{WW.}$} This is a large source of background. At $7$
   TeV, it has been measured by CMS~\cite{CMSww} and
   ATLAS~\cite{ATLASww} to be :
   $\sigma_{WW}=41.1\pm15.3\rm{(stat.)}\pm5.8\rm{(syst.)}\pm4.5\rm{(lumi.)}~\rm{pb}$
   and
   $\sigma_{WW}=41^{+20}_{-16}\rm{(stat.)}\pm5\rm{(syst.)}\pm1\rm{(lumi.)}~\rm{pb}$,
   with an integrated luminosity of 36 and 34 pb$^{-1}$,
   respectively. CMS collaboration also found~\cite{CMSwwzzwz} :
   $\sigma(pp\rightarrow WW +
   X)=55.3\pm3.3\rm{(stat.)}\pm6.9\rm{(syst.)}\pm3.3\rm{(lumi.)}~pb$. But
   pre-selection cuts (3 charged leptons out of which 2 have the same
   sign, 2 hard leptons) should reduce it to a negligible level.
 \item{$\mathbf {WZ.}$} The CMS collaboration
   measured~\cite{CMSwwzzwz} : $\sigma(pp\rightarrow WZ +
   X)=17.0\pm2.4\rm{(stat.)}\pm1.1\rm{(syst.)}\pm1.0\rm{(lumi.)}~pb$. This
   will give $\sim60$~fb for the final state cross section. A cut on
   the invariant mass of two leptons with opposite sign,
   $|M_Z-m_{ll}|>10~\rm{GeV}$ , can be applied in order to eliminate
   leptons coming from $Z$ decay. Moreover, if one considers leptons
   with different flavour, like for instance the channel
   $e^-\mu^+\mu^+$ + MET, this will be free from such a background.
 \item{$\mathbf{ZZ.}$} This channel is a background when one of the
   lepton is lost.  It has been measured at the LHC by the CMS collaboration~\cite{CMSwwzzwz} : $\sigma(pp\rightarrow ZZ + X)=3.8^{+1.5}_{-1.2}\rm{(stat.)}\pm0.2\rm{(syst.)}\pm0.2\rm{(lumi.)}~pb$. Again, cuts on
   the invariant mass of opposite signs leptons should allow to reduce
   it to a negligible level.
 \item{$\mathbf{t\overline{t}Z}$ and $\mathbf{b\overline{b}Z.}$} These
   constitute a background for final states involving jets. The
   production cross section is relatively large: $\sigma_{t\overline{t}Z}=205~\rm{fb}$ and $\sigma_{b\overline{b}Z}=50~\rm{pb}$. However, the cuts on
   the invariant mass of the leptons as well as $b$-tagging should
   reduce them to negligible levels.
 \item{$\mathbf{WWW.}$} Among the 3 gauge bosons background, this is
   the one with highest cross section. The production cross-section
   for three $W$ bosons is anyway lower than other background
   considered: $\sigma_{WWW}=71~\rm{fb}$, which becomes really
   negligible when the final state is considered.
\end{description}

All theses background sources can be reduces by cuts on the $p_T$ of
the leptons which are hard in the signal final state. Additional cuts
on number of jets or opposite-sign leptons' invariance mass can
further help to improve the signal over background ratio.

As it is clear, the aim of this section was just to describe the main
backgrounds affecting the considered signals. In order to give precise
estimation the entire simulation of the background should be
performed.


\subsection{Other relevant cases}

Even if we have discussed in details only the case of large mixing
with muons, there are other cases which can be relevant. Here we
briefly sketch their characteristics.

\begin{description}
\item[{\bf Mixing with electrons or taus.}] As already discussed in
  the literature~\cite{delAguila:2008cj}, the situation for mixing
  with electrons is similar to the one with muons and our analisis can
  be applied to it as well. On the other side, since detecting taus is
  more complicated, the discovery potential of channels involving taus
  is believed to be smaller.
\item[{\bf Mixing with 2 or 3 charged leptons.}] In such a case the
  triplet can couple to more than one family. The mixing angles are
  thus more constrained. As we have already shown (see
  Figs.~\ref{fig:BR_Vmu}-\ref{fig:BR_VeVmu}), the simultaneous
  presence of two (or three) non zero $V_\alpha$ would reduce the
  corresponding branching ratio by a small factor: if, for instance,
  two of them are taken to be equal, then the corresponding branching
  ratio will be decreased by a factor 2 with respect to the case with
  only one non-zero mixing angle (see
  Figs.~\ref{fig:BR_Vmu}-\ref{fig:BR_VeVmu}). However the pair
  production cross section of triplets is not affected by the mixing
  values and thus only the branching ratios and the mass of the
  triplet drive the relevant processes studied here.
\item[{\bf Small mixing angles, $\boldsymbol{\mathcal{O}(10^{-6})}$.}]
  This case is the ``most natural'' one, since here small neutrino
  masses can be accomodated without any cancellation or further source
  of suppression~\footnote{Notice that in this case the approximation
    of taking zero neutrino masses is no longer consistent and they
    should be turned on in the numerical simulations; for consistency
    also non-zero electron and muon masses should be considered, even
    if the effect of all these masses turns out to be
    negligible.}. Such small mixing angles drastically reduce the
  value of the triplet decay width, so that displaced vertexes up to
  few millimeters can be present (see also
  \cite{Franceschini:2008pz}). In case of finding an excess of events
  in some of the considered channels, the measurement of these
  displaced vertexes could be a clear signal that we are in presence
  of this kind of physics. The possible presence of a displaced vertex
  have to be taken into account when defining the reconstruction
  parameters for the data analisis (for example to reconstruct an
  interaction vertex). A detailed study of this topic is postponed to
  the analisis of real data.  A part from this, in general the cross
  sections are not affected and the analisis can proceed as in the
  case of large mixing.
\end{description}


\section{Conclusions}

In this paper we have described in details the minimal type III seesaw
model and its implementation in \texttt{FeynRules/MadGraph}. In
particular we have explicitly written all the couplings and we have
discussed the tests we have performed in order to validate the
implemented model. Even if the model has been tested only with
\texttt{MadGraph} which uses the unitary gauge, the Goldstone bosons
have been implemented as well, so that it can be used also with other
Monte Carlo generators such as
\texttt{CalcHep}~\cite{Christensen:2009jx}. As already stressed in the
Introduction, this is a necessary step to be done before proceeding to
the analysis of real LHC data.

In order to show an example of the utility of our model, we have
focused on a particular case --large mixing with muons,
$V_\mu=0.063$, and small triplet masses, 100~GeV, 120~GeV, 140~GeV--
and for these cases we have calculated the cross sections of the
relevant channels at the LHC running at 7~TeV. We have shown that several
events are expected for a luminosity of few fb$^{-1}$. We have
discussed the main background sources and the methods that can be
employed in order to reduce it. A more detailed study is beyond the
scope of this work, but, still at this level, we can expect that a
discovery at the LHC is possible, even in the 2011 run, if the mass of
the triplet is low enough and the background rejection is
good. Otherwise, in case of non-discovery, an upgrade of the bounds on
the triplet mass can be set.

\section*{Acknowledgments}

We gratefully thank Sara Vanini and Paolo Checchia for interesting
discussions and Claude Duhr for helping us in the publication of this
model on the FeynRules website.  C.B. thanks the Max Planck Institut
fuer Physik for the computing support and the project FPA2008-01430
for financial support.



\newpage

\appendix


\section{The explicit Lagrangian in the minimal model\label{app:couplings}}
\begin{eqnarray}
\textcolor{blue}{g^{CC}_{L}}&=&\left(\begin{array}{cc}
\left(1+\frac{V^T\wedge V}{2}\right)\mathcal{U}_{PMNS} & 
            -V^T\\
0 & \sqrt{2}\left(1-\frac{V\cdot V^T}{2}\right)\end{array}\right)\ = \ 
\nonumber
\end{eqnarray}
\begin{eqnarray}
=\left(\begin{array}{ccc}
\U_{e1}+\frac{\Ve V_\alpha}{2}\U_{\alpha 1} 
& \U_{e2}+\frac{\Ve V_{\alpha}}{2}\U_{\alpha 2} 
& \U_{e3}+\frac{\Ve V_{\alpha}}{2}\U_{\alpha 3}  \\
\U_{\mu 1}+\frac{\Vm V_{\alpha}}{2}\U_{\alpha 1} 
& \U_{\mu 2}+\frac{\Vm V_{\alpha}}{2}\U_{\alpha 2} 
& \U_{\mu 3}+\frac{\Vm V_{\alpha}}{2}\U_{\alpha 3}  \\
\U_{\tau 1}+\frac{\Vt V_\alpha}{2}\U_{\alpha 1} 
& \U_{\tau 2}+\frac{\Vt V_{\alpha}}{2}\U_{\alpha 2} 
& \U_{\tau 3}+\frac{\Vt V_{\alpha}}{2}\U_{\alpha 3}  \\
0 & 0 & 0 
 \end{array}\right|\nn\\
\left|\begin{array}{c}-\Ve \\ -\Vm \\ -\Vt \\ \sqrt{2}(1-\frac{\Ve^2+\Vm^2+\Vt^2}{2})\end{array}\right)\nonumber
\end{eqnarray}
\begin{eqnarray}
\textcolor{blue}{g^{CC}_{R}}&=& 
\left(\begin{array}{cc}
0 & -\sqrt{2}m_lV^T M_{\Sigma}^{-1}\\
-V\,\mathcal{U}_{PMNS}& 
             1-\frac{V\cdot V^T}{2}\end{array}\right)= \nonumber
\end{eqnarray}
\begin{eqnarray}
=\left(\begin{array}{cccc} 0 &0 &0 & -\sqrt{2}M_{\Sigma}^{-1} m_e \Ve \\
0 &0 &0 & -\sqrt{2}M_{\Sigma}^{-1} m_{\mu} \Vm \\
0 &0 &0 & -\sqrt{2}M_{\Sigma}^{-1} m_{\tau} \Vt \\
- V_\alpha \U_{\alpha 1} 
& - V_\alpha \U_{\alpha 2}
& - V_\alpha \U_{\alpha 3}
&  1-\frac{\Ve^2+\Vm^2+\Vt^2}{2} 
 \end{array}\right)\nonumber
\end{eqnarray}
\begin{eqnarray}
\textcolor{blue}{g^{NC}_{L}}&=&\left(\begin{array}{cc}
\frac{1}{2}-cos^2\theta_W-V^T\wedge V & \frac{1}{\sqrt{2}}V^T\\
\frac{1}{\sqrt{2}}V & V\cdot V^T -cos^2\theta_W\end{array}\right)=\nonumber
\end{eqnarray}
\begin{eqnarray}
=\left(\begin{array}{cccc} \frac{1}{2}-cos^2\theta_W-\Ve^2 & \Ve\Vm & \Ve\Vt & \frac{\Ve}{\sqrt{2}}\\
\Ve\Vm & \frac{1}{2}-cos^2\theta_W-\Vm^2 & \Vm\Vt & \frac{\Vm}{\sqrt{2}}\\
\Ve\Vt & \Vm\Vt &  \frac{1}{2}-cos^2\theta_W-\Vt^2 & \frac{\Vt}{\sqrt{2}}\\
 \frac{\Ve}{\sqrt{2}} &  \frac{\Vm}{\sqrt{2}} &  \frac{\Vt}{\sqrt{2}} &\Ve^2+\Vm^2+\Vt^2-cos^2\theta_W\end{array}\right)\nn
\end{eqnarray}
\begin{eqnarray}
\textcolor{blue}{g^{NC}_{R}}&=& 
\left(\begin{array}{cc}
1-cos^2\theta_W & \sqrt{2}m_lV^T M_{\Sigma}^{-1}\\
\sqrt{2}M_{\Sigma}^{-1}Vm_l& -cos^2\theta_W\end{array}\right)= \nonumber
\end{eqnarray}
\begin{eqnarray}
=\left(\begin{array}{cccc} 1-cos^2\theta_W & 0 & 0 & \sqrt{2}M_{\Sigma}^{-1}m_e\Ve\\
 0 & 1-cos^2\theta_W & 0 & \sqrt{2}M_{\Sigma}^{-1}m_{\mu}\Vm\\
 0 & 0 & 1-cos^2\theta_W  & \sqrt{2}M_{\Sigma}^{-1}m_{\tau}\Vt\\
\sqrt{2}M_{\Sigma}^{-1}m_e\Ve & \sqrt{2}M_{\Sigma}^{-1}m_{\mu}\Vm & \sqrt{2}M_{\Sigma}^{-1}m_{\tau}\Vt & -cos^2\theta_W\end{array}\right) \nonumber
\end{eqnarray}
\begin{eqnarray}
\textcolor{blue}{g^{NC}_{\nu}}&=& 
\left(\begin{array}{cc}
1-\mathcal{U}_{PMNS}^T\,V^T\wedge V\,\mathcal{U}_{PMNS}&\mathcal{U}_{PMNS}^T V^T \\
V\,\mathcal{U}_{PMNS}& V\cdot V^T \end{array}\right)=\nonumber
\end{eqnarray}
\begin{eqnarray}
=\left(\begin{array}{cc}
1- \U_{\alpha 1} V_\alpha V_\beta \U_{\beta 1} 
& -\U_{\alpha 1} V_\alpha V_\beta \U_{\beta 2} \\
- \U_{\alpha 2} V_\alpha V_\beta \U_{\beta 1} 
& 1 -\U_{\alpha 2} V_\alpha V_\beta \U_{\beta 2} \\
- \U_{\alpha 3} V_\alpha V_\beta \U_{\beta 1} 
& -\U_{\alpha 3} V_\alpha V_\beta \U_{\beta 2} \\
V_\alpha \U_{\alpha 1} 
& V_\alpha \U_{\alpha 2} \\
\end{array}\right|\nn\\
\left|\begin{array}{cc} 
 - \U_{\alpha 1} V_\alpha V_\beta \U_{\beta 3} 
&V_\alpha \U_{\alpha 1} \\
 - \U_{\alpha 2} V_\alpha V_\beta \U_{\beta 3}
&V_\alpha \U_{\alpha 2} \\
 1 - \U_{\alpha 3} V_\alpha V_\beta \U_{\beta 3}
&V_\alpha \U_{\alpha 3}\\
 V_\alpha \U_{\alpha 3} 
& \Ve^2+\Vm^2+\Vt^2
\end{array}\right)\nonumber
\end{eqnarray}
\begin{eqnarray}
\textcolor{blue}{g^{H\ell}_{L}}&=& 
\left(\begin{array}{cc}
\frac{m_l}{v}\left(1-3V^T\wedge V\right) &\sqrt{2} \frac{m_l}{v}V^T\\
\sqrt{2}\frac{M_{\Sigma}}{v}V\cdot(1-V^T\wedge V)+\sqrt{2}M_{\Sigma}^{-1}V\frac{m_l^2}{v} & 2\frac{M_{\Sigma}}{v} V\cdot V^T\end{array}\right)=\nonumber
\end{eqnarray}
\begin{eqnarray}
=\left(\begin{array}{cc}
\frac{m_e}{v}(1-3\Ve^2) & -3\frac{m_e}{v}\Ve\Vm \\
-3\frac{m_\mu}{v}\Vm\Ve & \frac{m_\mu}{v}(1-3\Vm^2) \\
-3\frac{m_{\tau}}{v}\Vt\Ve & -3\frac{m_{\tau}}{v}\Vt\Vm \\
\sqrt{2}\frac{M_{\Sigma}}{v}\Ve(1-\Ve^2-\Vm^2-\Vt^2)+\sqrt{2}M_{\Sigma}^{-1}\frac{m_e^2}{v}\Ve & \sqrt{2}\frac{M_{\Sigma}}{v}\Vm(1-\Ve^2-\Vm^2-\Vt^2)+\sqrt{2}M_{\Sigma}^{-1}\frac{m_{\mu}^2}{v}\Vm
\end{array}\right|\nn\\
\left|\begin{array}{cc}-3\frac{m_e}{v}\Ve\Vt & \sqrt{2}\frac{m_e}{v}\Ve \\
 -3\frac{m_\mu}{v}\Vm\Vt & \sqrt{2}\frac{m_\mu}{v}\Vm\\
\frac{m_{\tau}}{v}(1-3\Vt^2) & \sqrt{2}\frac{m_\tau}{v}\Vt\\
\sqrt{2}\frac{M_{\Sigma}}{v}\Vt(1-\Ve^2-\Vm^2-\Vt^2)+\sqrt{2}M_{\Sigma}^{-1}\frac{m_{\tau}^2}{v}\Vt & 2\frac{M_{\Sigma}}{v}(\Ve^2+\Vm^2+\Vt^2) \end{array}\right)\nonumber
\end{eqnarray}
\begin{eqnarray}
\textcolor{blue}{g^{H\nu}_{L}}&=& 
\left(\begin{array}{cc}
\frac{\sqrt{2}}{v} m_\nu^d
&\frac{\sqrt{2}}{v}  m_\nu^d \mathcal{U}_{PMNS}^T  V^T\\
\frac{\sqrt{2}}{v}
\left(1-\epsilon^\prime\right) M_\Sigma V \mathcal{U}_{PMNS}
&\frac{\sqrt{2}}{v} M_\Sigma \epsilon^\prime \end{array}\right)=\nonumber
\end{eqnarray}
\begin{eqnarray}
= \frac{\sqrt{2}}{v}
\left(\begin{array}{cccc}
m_{\nu1}&0&0&m_{\nu1} {\mathcal{U}_{PMNS}}_{\alpha 1} V_\alpha\\
0&m_{\nu2}&0&m_{\nu2} {\mathcal{U}_{PMNS}}_{\alpha 2} V_\alpha\\
0&0&m_{\nu3}&m_{\nu3} {\mathcal{U}_{PMNS}}_{\alpha 3} V_\alpha\\
\left( 1- \epsilon^\prime\right) M_\Sigma  V_\alpha
{\mathcal{U}_{PMNS}}_{\alpha 1}
& \left( 1- \epsilon^\prime\right)  M_\Sigma  V_\alpha
{\mathcal{U}_{PMNS}}_{\alpha 2}
& \left( 1- \epsilon^\prime\right)  M_\Sigma  V_\alpha
{\mathcal{U}_{PMNS}}_{\alpha 3}
&M_\Sigma \epsilon^\prime
\end{array}\right)\nonumber
\end{eqnarray}
\begin{eqnarray}
\textcolor{blue}{g^{\eta\ell}_{L}}&=&
\left(\begin{array}{cc}-\frac{m_l}{v}(1+V^T\wedge V) 
& -\frac{m_l}{v} \sqrt{2} V^T \\
\frac{M_\Sigma}{v} \sqrt{2} V 
(1- V^T\wedge V - \frac{m^2_l}{M_\Sigma^2})
& \frac{2 M_\Sigma}{v} V\cdot V^T \end{array}\right)=\nonumber
\end{eqnarray}
\begin{eqnarray}
=\left(\begin{array}{cc}
-\frac{m_e}{v}(1+\Ve^2) & -\frac{m_e}{v}\Ve\Vm \\
-\frac{m_\mu}{v}\Vm\Ve & -\frac{m_\mu}{v}(1+\Vm^2) \\
-\frac{m_{\tau}}{v}\Vt\Ve & -\frac{m_{\tau}}{v}\Vt\Vm \\
\sqrt{2}\frac{M_{\Sigma}}{v}\Ve(1-\Ve^2-\Vm^2-\Vt^2)-\sqrt{2}M_{\Sigma}^{-1}\frac{m_e^2}{v}\Ve & \sqrt{2}\frac{M_{\Sigma}}{v}\Vm(1-\Ve^2-\Vm^2-\Vt^2)-\sqrt{2}M_{\Sigma}^{-1}\frac{m_{\mu}^2}{v}\Vm
\end{array}\right|\nn\\
\left|\begin{array}{cc}-\frac{m_e}{v}\Ve\Vt & -\sqrt{2}\frac{m_e}{v}\Ve \\
 -\frac{m_\mu}{v}\Vm\Vt & -\sqrt{2}\frac{m_\mu}{v}\Vm\\
-\frac{m_{\tau}}{v}(1+\Vt^2) & -\sqrt{2}\frac{m_\tau}{v}\Vt\\
\sqrt{2}\frac{M_{\Sigma}}{v}\Vt(1-\Ve^2-\Vm^2-\Vt^2)-\sqrt{2}M_{\Sigma}^{-1}\frac{m_{\tau}^2}{v}\Vt & 2\frac{M_{\Sigma}}{v}(\Ve^2+\Vm^2+\Vt^2) \end{array}\right)\nonumber
\end{eqnarray}
\begin{eqnarray}
\textcolor{blue}{g_L^{\phi}}=\left(\begin{array}{cc} 
\sqrt{2}\frac{m_l}{v}(1-\frac{V^T\wedge V}{2})\mathcal{U}_{PMNS} 
& \frac{\sqrt{2}}{v}m_l V^T\\
2 M_{\Sigma}^{-1}V\frac{ m_l^2}{v}\mathcal{U}_{PMNS} 
& 0\end{array}\right)=\nonumber
\end{eqnarray}
\begin{eqnarray}
=\left(\begin{array}{cccc} 
\frac{\sqrt{2}}{v}(m_e(\delta_{e\alpha}-\frac{V_eV_\alpha}{2})\mathcal{U}_{PMNS_{\alpha 1}})  
&   \frac{\sqrt{2}}{v}(m_e(\delta_{e\alpha}-\frac{V_eV_\alpha}{2})\mathcal{U}_{PMNS_{\alpha 2}})  
&   \frac{\sqrt{2}}{v}(m_e(\delta_{e\alpha}-\frac{V_eV_\alpha}{2})\mathcal{U}_{PMNS_{\alpha 3}})  
&   \frac{\sqrt{2}}{v}m_eV_e\\
\frac{\sqrt{2}}{v}(m_\mu(\delta_{\mu\alpha}-\frac{V_\mu V_\alpha}{2})\mathcal{U}_{PMNS_{\alpha 1}})  
&   \frac{\sqrt{2}}{v}(m_\mu(\delta_{\mu\alpha}-\frac{V_\mu V_\alpha}{2})\mathcal{U}_{PMNS_{\alpha 2}})  
&   \frac{\sqrt{2}}{v}(m_\mu(\delta_{\mu\alpha}-\frac{V_\mu V_\alpha}{2})\mathcal{U}_{PMNS_{\alpha 3}})  
&   \frac{\sqrt{2}}{v}m_\mu V_\mu\\
\frac{\sqrt{2}}{v}(m_\tau(\delta_{\tau\alpha}-\frac{V_\tau V_\alpha}{2})\mathcal{U}_{PMNS_{\alpha 1}})  
&   \frac{\sqrt{2}}{v}(m_\tau(\delta_{\tau\alpha}-\frac{V_\tau V_\alpha}{2})\mathcal{U}_{PMNS_{\alpha 2}})  
&   \frac{\sqrt{2}}{v}(m_\tau(\delta_{\tau\alpha}-\frac{V_\tau V_\alpha}{2})\mathcal{U}_{PMNS_{\alpha 3}})  
&   \frac{\sqrt{2}}{v}m_\tau V_\tau\\
2M^{-1}_{\Sigma}V_\alpha \frac{m_{\alpha}^2}{v} \mathcal{U}_{PMNS_{\alpha 1}} & 2M^{-1}_{\Sigma}V_\alpha \frac{m_{\alpha}^2}{v} \mathcal{U}_{PMNS_{\alpha 2}} & 2M^{-1}_{\Sigma}V_\alpha \frac{m_{\alpha}^2}{v} \mathcal{U}_{PMNS_{\alpha 3}}  &  0\end{array}\right)\nonumber
\end{eqnarray}
\begin{eqnarray}
\textcolor{blue}{g_R^{\phi}}=\left(\begin{array}{cc} 
-\sqrt{2}\mathcal{U}_{PMNS}\frac{m_{\nu}^d}{v} 
& (V^T-(V^T\wedge V)\cdot V^T-V^T\cdot\frac{V.V^T}{2})
\sqrt{2}\frac{M_\Sigma}{v}
-2\sqrt{2} \mathcal{U}_{PMNS}\frac{m_{\nu}^d}{v} \mathcal{U}_{PMNS}^{T}V^T\\
-2\frac{M_{\Sigma}}{v}V(1-\frac{(V^T\wedge V)}{2})\mathcal{U}_{PMNS}  
& 0 
\end{array}\right)=\nonumber
\end{eqnarray}
\begin{eqnarray}
=\left(\begin{array}{cccc}
-\frac{\sqrt{2}}{v}m_{\nu_1}\mathcal{U}_{e1}  
&-\frac{\sqrt{2}}{v}  m_{\nu_2}\mathcal{U}_{e2}  
& -\frac{\sqrt{2}}{v}  m_{\nu_3}\mathcal{U}_{e3}    \\
-\frac{\sqrt{2}}{v}m_{\nu_1}\mathcal{U}_{\mu 1}  
&-\frac{\sqrt{2}}{v}  m_{\nu_2}\mathcal{U}_{\mu 2}  
& -\frac{\sqrt{2}}{v}  m_{\nu_3}\mathcal{U}_{\mu 3}   \\
-\frac{\sqrt{2}}{v}m_{\nu_1}\mathcal{U}_{\tau 1}  
&-\frac{\sqrt{2}}{v}   m_{\nu_2}\mathcal{U}_{\tau 2}  
& -\frac{\sqrt{2}}{v}  m_{\nu_3}\mathcal{U}_{\tau 3}   \\
\frac{M_{\Sigma}}{v}  V_{\alpha}(V_e^2+V_\mu^2+V_\tau^2-2) \mathcal{U}_{\alpha 1}  
&\frac{M_{\Sigma}}{v} V_{\alpha}(V_e^2+V_\mu^2+V_\tau^2-2) \mathcal{U}_{\alpha 2} 
&\frac{M_{\Sigma}}{v} V_{\alpha}(V_e^2+V_\mu^2+V_\tau^2-2) \mathcal{U}_{\alpha 3} 
\end{array}\right|\nonumber\\
\nonumber\\
\left|\begin{array}{c}
\sqrt{2}\frac{M_{\Sigma}}{v}V_e(1-\frac{3}{2}(V_e^2+V_\mu^2+V_\tau^2))
-2\sqrt{2}\frac{m_{\nu_i}}{v}\mathcal{U}_{ei} \mathcal{U}_{\alpha i}V_{\alpha}\\
\sqrt{2}\frac{M_{\Sigma}}{v}V_\mu(1-\frac{3}{2}(V_e^2+V_\mu^2+V_\tau^2))
-2\sqrt{2}\frac{m_{\nu_i}}{v}\mathcal{U}_{\mu i}\mathcal{U}_{\alpha i}V_{\alpha}\\
\sqrt{2}\frac{M_{\Sigma}}{v}V_\tau(1-\frac{3}{2}(V_e^2+V_\mu^2+V_\tau^2))
-2\sqrt{2}\frac{m_{\nu_i}}{v}\mathcal{U}_{\tau i}\mathcal{U}_{\alpha i}V_{\alpha}\\
0\end{array}\right)\nonumber
\end{eqnarray}

In the above expressions repetead flavour indexes are summed. As we
will discuss later, we will take neutrino masses equal to zero, except
in the case of small mixing angles~\footnote{In this case, indeed, for
  consistency we will turn neutrino masses, as well as electron and
  muon masses, on. However, this will not basically affect the
  result.}.

\newpage
\section{Tables for the validation of the implementation\label{app:validation}}

\begin{table}[!h]
\begin{center}
\begin{tabular}{|l|l|l|l|}
\hline
Process & \texttt{sm\_FR} & \texttt{typeIIIseesaw1\_MG} & comparison \\
\hline
\hline
top decay & 1.53174916 & 1.55409729 & 1.45899\% \\
\hline
W decay &  2.00335798 & 2.00322925 & 0.00642571\% \\
\hline
Z decay & 2.41539342 & 2.41481975 & 0.0237506\% \\
\hline
BR(w+ $\to$ v e+ ) & 1.11025062e-01 & 1.11142e-01 & 0.105326\%\\
\hline
BR(w+ $\to$ v m+ )& 1.11036355e-01 & 1.11331e-01 & 0.265359\%\\
\hline
BR(w+ $\to$ v tt+ ) & 1.12013868e-01 & 1.11018e-01 & 0.8962\%\\
\hline
BR(w+ $\to$ c d~ ) & 1.69615944e-02 &  1.69574065e-02 & 0.0246905\%\\
\hline
BR(w+ $\to$ u d~ ) & 3.14853587e-01 & 3.16304871e-01 & 0.460939\% \\
\hline
BR(w+ $\to$ c s~ ) & 3.17238100e-01 & 3.16278512e-01 & 0.302482\% \\
\hline
BR(w+ $\to$ u s~ ) & 1.68714343e-02 & 1.69683505e-02 & 0.574441\% \\
\hline
BR(z $\to$ e- e+ ) &  3.45878542e-02 & 3.45049797e-02 & 0.239606\%\\
\hline
BR(z $\to$ m- m+ ) & 3.46182266e-02 & 3.49703234e-02 & 1.01709\%\\
\hline
BR(z $\to$ tt- tt+ ) & 3.45433552e-02 & 3.45770661e-02 & 0.0975901\%\\
\hline
BR(z $\to$ invisible) & 0.205237 & 0.205557 & 0.155917\% \\
\hline
BR(z $\to$ b b~ ) & 1.51238258e-01 & 1.50200176e-01 & 0.686388\%\\
\hline
BR(z $\to$ c c~ ) & 1.17361782e-01 & 1.17167722e-01 & 0.165352\%\\
\hline
BR(z $\to$ d d~ ) & 1.52782011e-01 & 1.52925551e-01 & 0.0939509\%\\
\hline
BR(z $\to$ s s~ ) & 1.52615959e-01 & 1.51787006e-01 & 0.543163\%\\
\hline
BR(z $\to$ u u~ ) & 1.17015696e-01 &  1.18309630e-01 & 0.10578\%\\
\hline 
\end{tabular}
\end{center}
\caption{Comparison of decay widths and branching ratios between the
  model \texttt{sm\_FR} and \texttt{typeIIIseesaw1\_MG}.}
\label{tab:validation-Br}
\end{table}

\begin{table}[!h]
\begin{center}
\begin{tabular}{|l|c|c|c|}
\hline
Process & \texttt{sm\_FR} & \texttt{typeIIIseesaw1\_MG} & comparison \\
\hline
\hline
$e^+e^-\rightarrow e^+e^-$ & $7.457\rm{e}{+2}$ & $7.450\rm{e}{+2}$ & 0.095 \% \\
$e^+e^-\rightarrow \mu^+\mu^-$ & $1.125\rm{e}{-1}$ & $1.126\rm{e}{-1}$ & 0.09 \%\\
$e^+e^-\rightarrow \nu^+\nu^-$ & $5.185\rm{e}{+1}$ & $5.180\rm{e}{+1}$ & 0.10\%\\
\hline
$\tau^+\tau^-\rightarrow W^+ W^-$ & $2.629\rm{e}{+0}$ & $2.625\rm{e}{+0}$ & 0.15\%\\
$\tau^+\tau^-\rightarrow ZZ$ & $1.448\rm{e}{-1}$ &  $1.449\rm{e}{-1}$ & 0.07\% \\
$\tau^+\tau^-\rightarrow Z\gamma$ & $7.208\rm{e}{-1}$ & $7.219\rm{e}{-1}$ & 0.15\% \\
$\tau^+\tau^-\rightarrow \gamma\gamma$ & $1.020\rm{e}{+0}$ & $1.020\rm{e}{+0}$ & -- \\
\hline
$ZZ\rightarrow ZZ$ & $5.997\rm{e}{-1}$ & $5.996\rm{e}{-1}$ &0.017\%\\
$W^+ W^-\rightarrow ZZ$ & $2.996\rm{e}{+2}$ & $2.995\rm{e}{+2}$ &0.033\%\\
$HH\rightarrow ZZ$ & $6.763\rm{e}{+1}$ & $6.763\rm{e}{+1}$ & --\\
$HH\rightarrow W^+ W^-$ & $1.046\rm{e}{+2}$ & $1.039\rm{e}{+2}$ & 0.57\%\\
\hline
$GG\rightarrow GG$ & $3.084\rm{e}{+5}$ & $3.079\rm{e}{+5}$ & 0.16\% \\
\hline
$u\overline{u}\rightarrow GG$ & $1.981\rm{e}{+2}$ & $1.980\rm{e}{+2}$ &0.05\% \\
$u\overline{u}\rightarrow W^+W^-$ & $8.711\rm{e}{-1}$ & $8.720\rm{e}{-1}$ & 0.10\% \\
$u\overline{u}\rightarrow ZZ$ & $8.783\rm{e}{-2}$ & $8.800\rm{e}{-2}$ & 0.19\%\\
$u\overline{u}\rightarrow Z\gamma$ & $1.215\rm{e}{-1}$ & $1.216\rm{e}{-1}$ & 0.08\% \\
$u\overline{u}\rightarrow \gamma\gamma$ & $6.725\rm{e}{-2}$ & $6.714\rm{e}{-2}$ & 0.13\%\\
$u\overline{u}\rightarrow s\overline{s}$ & $7.809\rm{e}{+0}$ & $7.807\rm{e}{+0}$ &0.026 \% \\
$u\overline{d}\rightarrow c\overline{s}$ & $1.040\rm{e}{-1}$ & $1.040\rm{e}{-1}$ & --\\
$u\overline{s}\rightarrow c\overline{d}$ & $3.000\rm{e}{-4}$ & $2.999\rm{e}{-4}$ & 0.033\%\\
\hline
$t\overline{t}\rightarrow GG$ & $7.352\rm{e}{+1}$ & $7.349\rm{e}{+1}$ & 0.027\%\\
$t\overline{t}\rightarrow W^+W^-$ & $7.521\rm{e}{+0}$ & $7.512\rm{e}{+0}$ & 0.12\% \\
$t\overline{t}\rightarrow ZZ$ & $7.875\rm{e}{-1}$ & $7.899\rm{e}{-1}$ & 0.30\%\\
$t\overline{t}\rightarrow Z\gamma$ & $4.778\rm{e}{-1}$ & $4.771\rm{e}{-1}$ & 0.15\% \\
$t\overline{t}\rightarrow \gamma\gamma$ & $3.096\rm{e}{-2}$ & $3.091\rm{e}{-2}$ & 0.161\%\\
$t\overline{t}\rightarrow u\overline{u}$ & $3.139\rm{e}{+0}$ & $3.130\rm{e}{+0}$ & 0.28\%\\
\hline
\end{tabular}
\end{center}
\caption{Selection of $2\to 2$ processes. The \texttt{FeynRules}
  generated Standard Model implementations in
  \texttt{MadGraph/MadEvent} is denoted \texttt{sm\_FR} and the one of
  the type III Seesaw is denoted \texttt{typeIIIseesaw1\_MG}. The
  center-of-mass energy is fixed to 1~TeV and a $p_T$ cut of 20~GeV is
  applied to each final state particle. }
\label{tab:validation-2to2}
\end{table}

\begin{table}[!h]
\begin{center}
\begin{tabular}{|l|l|l|l|}
\hline
$M_\Sigma$ & $\sigma(pp\to \Sigma^+ \Sigma^0)(fb)$ 
& $\sigma(pp\to \Sigma^+ \Sigma^-)(fb)$ & $\sigma(pp\to \Sigma^- \Sigma^0)(fb)$ \\
\hline
100 & $1.126\rm{e}{+4}$ & $9.125\rm{e}{+3}$ & $6.914\rm{e}{+3}$ \\
\hline
120 & $5.818\rm{e}{+3}$ & $4.673\rm{e}{+3}$ & $3.480\rm{e}{+3}$  \\
\hline
140 & $3.373\rm{e}{+3}$ & $2.673\rm{e}{+3}$ & $1.957\rm{e}{+3}$  \\
\hline
160 & $2.100\rm{e}{+3}$ & $1.646\rm{e}{+3}$ & $1.184\rm{e}{+3}$ \\
\hline
180 & $1.382\rm{e}{+3}$ & $1.071\rm{e}{+3}$ & $7.604\rm{e}{+2}$ \\
\hline
200 & $9.471\rm{e}{+2}$ & $7.273\rm{e}{+2}$ & $5.073\rm{e}{+2}$ \\
\hline
300 & $2.136\rm{e}{+2}$ & $1.564\rm{e}{+2}$ & $1.023\rm{e}{+2}$ \\
\hline
400 & $7.012\rm{e}{+1}$ & $4.847\rm{e}{+1}$ & $3.039\rm{e}{+1}$ \\
\hline
600 & $1.280\rm{e}{+1}$ & $8.307$ & $4.713$ \\
\hline
800 & $3.290$ & 1.993 & 1.068 \\
\hline
1000 & $1.018$ & $5.896\rm{e}{-1}$ & $2.978\rm{e}{-1}$ \\
\hline 
\end{tabular}
\end{center}
\caption{Production cross sections at 14~TeV. These values have been
  obtained with \texttt{MadGraph/MadEvent} and the acceptance cuts
  implemented are listed in Table~\ref{tab:acceptance-cuts}.
  Fig.~\ref{fig:prod-14TeV} shows the interpolated curves.}
\label{tab:prod-14TeV}
\end{table}

\begin{figure}[!h]
\begin{center}
\epsfig{figure=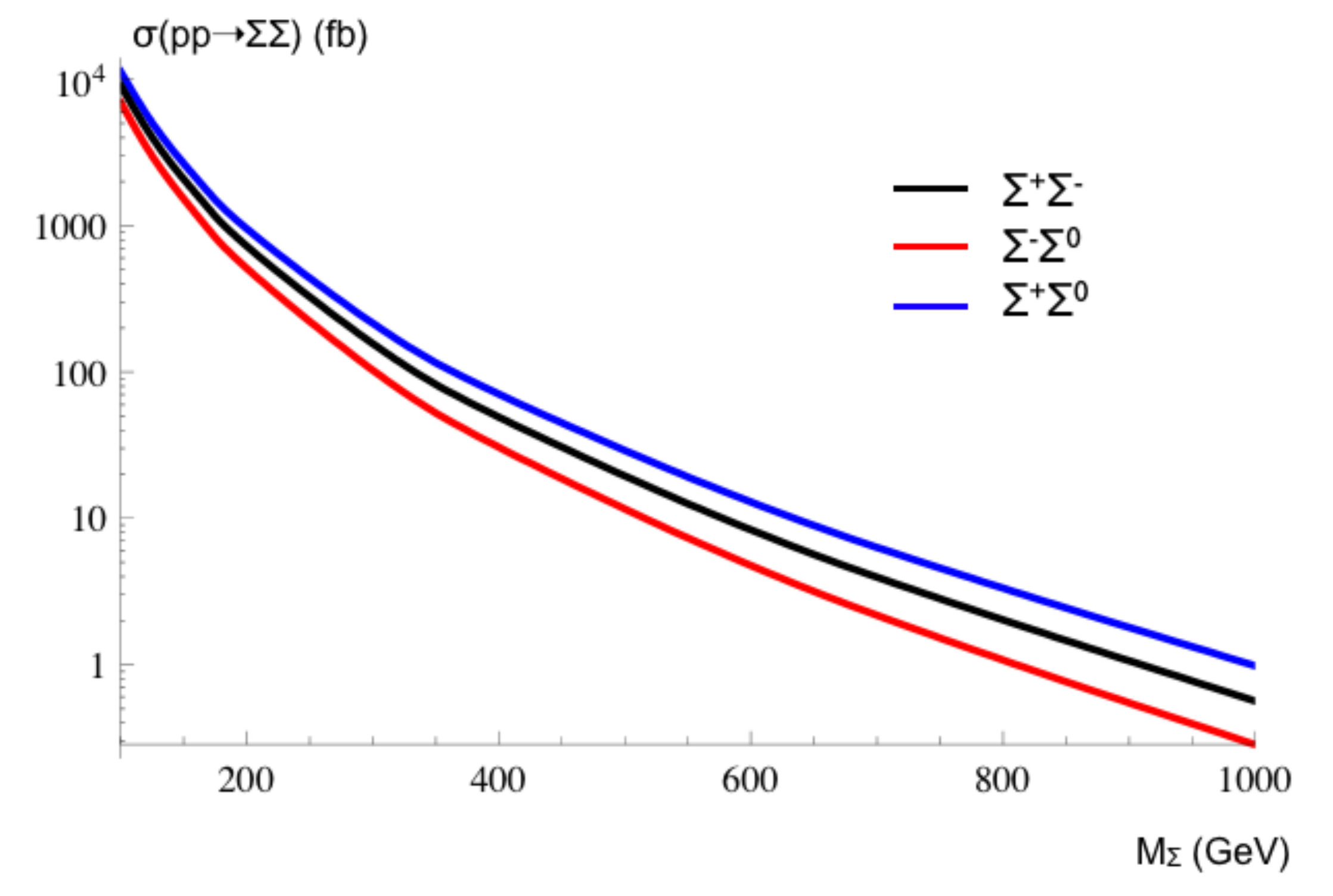,width=13cm}
\caption{\label{fig:prod-14TeV} Production of a pair of triplets at
  14~TeV at the LHC. The mixing parameters as been set to $V_\mu =
  0.063$ and $V_e=V_\tau=0$.}
\label{fig:Br1}
\end{center}
\end{figure}

\newpage
~
\newpage
~\newpage

\section{Cross sections of the relevant channels at 7~TeV}
\label{app:cross}

\begin{table}[!h]
\begin{center}
\begin{tabular}{|l|r|r|r|l|r|r|r|}
\hline
 Process  
& \multicolumn{3}{|c|}{Cross Sections (fb)}
& Final State          
& \multicolumn{3}{|c|}{\textcolor{blue}{Final State Cross Section (fb)}} \\
\hline 
  & 100 GeV & 120 GeV & 140 GeV &  & \textcolor{blue}{100 GeV} & \textcolor{blue}{120 GeV} & \textcolor{blue}{140 GeV}  \\ 
\hline 
\multicolumn{8}{|c|}{\textcolor{blue}{Final State $++$}} \\
\hline
 $ W^- \mu^+ Z \mu^+ $ & $2.36\rm{e}+2$ & $2.02\rm{e}+2$ & $1.16\rm{e}+2$ 
& $\mu^{+} \mu^{+} hadr$ & $108$ & $92.7$ & $53.4$ \\ 
\hline
&  & & 
& $\mu^{+} \mu^{+} \nu\nu hadr$ & $32.4$ & $27.8$ & $15.9$ \\ 
\hline
 $ W^- \mu^+ W^+ \nu $ & $1.66\rm{e}+3$ & $6.06\rm{e}+2$ & $2.82\rm{e}+2$ 
& $\mu^{+} \mu^{+} \nu\nu hadr$ & $124$ & $45.3$ & $21.1$ \\ 
\hline
 $ W^- \mu^+  h \mu^+ $ & $1.22\rm{e}-3$ & $1.39\rm{e}-1$ & $1.40\rm{e}+1$
& $\mu^{+} \mu^{+} hadr$ & - & - & $8.9$ \\ 
\hline
&  & & 
& $\mu^{+} \mu^{+} \nu\nu hadr$ &- &- & - \\ 
\hline
\multicolumn{5}{|c|}{\textcolor{blue}{Total Cross Sections $\mu^{+} \mu^{+}$ + jets + missing $E_T$}} & 156.4 & 73.1 & 37.0 \\ 
\hline
\multicolumn{5}{|c|}{\textcolor{blue}{Total Cross Sections $\mu^{+} \mu^{+}$ + jets}} & 108 & 92.7 & 62.3 \\ 
\hline
\multicolumn{8}{|c|}{\textcolor{blue}{Final State $--$}} \\
\hline
 $  W^+ \mu^- Z \mu^- $ & $1.27\rm{e}+2$ & $1.04\rm{e}+2$ & $5.67\rm{e}+1$ 
& $\mu^{-} \mu^{-} hadr$ & $58.3$ & $47.7$ & $26.1$ \\ 
\hline
 &  & &
& $\mu^{-} \mu^{-} \nu \nu hadr$ & $17.4$ & $14.3$ & $7.8$ \\ 
\hline
 $  W^+ \mu^- W^- \nu $ & $8.94\rm{e}+2$ & $3.11\rm{e}+2$ & $1.39\rm{e}+2$ 
& $\mu^{-} \mu^{-} \nu \nu hadr$ & $67.0$ & $23.3$ & $10.4$ \\ 
\hline
 $ W^+ \mu^-  h \mu^- $ & $5.87\rm{e}-6$ & $7.13\rm{e}-2$ & $6.86$
& $\mu^{-} \mu^{-} hadr$ & - & - & $4.4$\\ 
\hline
& & &
& $\mu^{-} \mu^{-} \nu \nu hadr$ & - & - & - \\ 
\hline
\multicolumn{5}{|c|}{\textcolor{blue}{Total Cross Sections $\mu^{-} \mu^{-}$ + jets + missing $E_T$}} & 84.4 & 37.6 & 18.2\\ 
\hline
\multicolumn{5}{|c|}{\textcolor{blue}{Total Cross Sections $\mu^{-} \mu^{-}$ + jets}} & 58.3 & 47.7 & 30.5\\ 
\hline
\end{tabular}
\end{center}
\caption{Final states with two muons of the same sign for
  $V_e=V_\tau=0$, $V_\mu=0.063$. The final cross sections have been
  computed using the measured branching ratios, except for the Higgs,
  whose branching ratios have been calculated assuming a mass of
  120~GeV. Only channels with a final cross section higher than 0.1
  have been reported.}
\label{tab:dileptons}
\end{table}

\begin{table}[!h]
\begin{center}
\begin{tabular}{|l|r|r|r|l|r|r|r|}
\hline
 Process  
& \multicolumn{3}{|c|}{Cross Sections (fb)}
& Final State          
& \multicolumn{3}{|c|}{\textcolor{blue}{Final State Cross Section (fb)}} \\
\hline 
 & 100 GeV & 120 GeV & 140 GeV &  & \textcolor{blue}{100 GeV} & \textcolor{blue}{120 GeV} & \textcolor{blue}{140 GeV}  \\ 
\hline 
\hline 
\multicolumn{8}{|c|}{\textcolor{blue}{Final State $+ + -$}} \\
\hline
 $ W^+ \mu^-  W^+ \nu $  & $1.66\rm{e}+3$  & $6.08\rm{e}+2$  & $2.82\rm{e}+2$   
& $\mu^{+} \mu^{+} \mu^{-} \nu \nu \nu$ & $20.9$ & $7.7$ &  $3.5$ \\ 
\hline
 $ W^- \mu^+  W^+ \nu $  & $1.66\rm{e}+3$  & $6.06\rm{e}+2$  & $2.82\rm{e}+2$   
& $\mu^{+} \mu^{+} \mu^{-} \nu \nu \nu$ & $20.9$ & $7.7$ &  $3.5$ \\ 
\hline
 $ W^+ \mu^- Z \mu^+ $ & $2.36\rm{e}+2$ & $2.03\rm{e}+2$ & $1.16\rm{e}+2$  
& $\mu^{+} \mu^{+} \mu^{-} \nu hadr$ & $18.2$ & $15.7$ & $8.9$ \\ 
\hline
   &   &  &  
& $\mu^{+} \mu^{+} \mu^{-} \nu \nu \nu $ & $5.5$ & $4.7$ & $2.7$\\ 
\hline
 $ W^- \mu^+ Z \mu^+ $ & $2.36\rm{e}+2$ & $2.02\rm{e}+2$ & $1.16\rm{e}+2$  
& $\mu^{+} \mu^{+} \mu^{-} \nu hadr$ & $18.3$ & $15.6$ & $8.9$ \\ 
\hline
  &   &  &  
& $\mu^{+} \mu^{+} \mu^{-} \nu \nu \nu $ & $5.5$ & $4.6$ & $2.6$ \\ 
\hline
 $ W^+ \nu Z \nu $ & $4.62\rm{e}+2$  & $4.02\rm{e}+2$ & $2.32\rm{e}+2$
& $\mu^{+} \mu^{+} \mu^{-} \nu \nu \nu $ & $1.8$ & $1.6$ & $0.9$ \\ 
\hline
 $ Z \mu^+ Z \nu $ & $6.55\rm{e}+1$ & $1.35\rm{e}+2$ & $9.48\rm{e}+1$  
& $\mu^{+} \mu^{+} \mu^{-} \nu hadr $ & $1.6$  & $3.2$ & $2.3$ \\ 
\hline
   &   &  &  
& $\mu^{+} \mu^{+} \mu^{-} \nu \nu \nu $ & $0.47$ & $0.98$ & $0.68$ \\ 
\hline
 $ Z \mu^+ h \nu $  & $6.80\rm{e}-4$  & $1.54\rm{e}-1$ & $2.28\rm{e}+1$  
& $\mu^{+} \mu^{+} \mu^{-} \nu hadr $ & - & - & 0.76 \\ 
\hline
 $ W^- \nu Z \mu^+ $ & $3.61\rm{e}+2$ & $3.08\rm{e}+2$ & $1.71\rm{e}+2$
& $\mu^{+} \mu^{+} \mu^{-} \nu hadr$ & $8.4$ & $7.2$ & $4.0$ \\ 
\hline
 $ W^+ \mu^- h \mu^+ $  & $1.22\rm{e}-3$ & $1.39\rm{e}-1$ & $1.40\rm{e}+1$   
& $\mu^{+} \mu^{+} \mu^{-} \nu hadr$ & - & - & $1.5$\\ 
\hline
 $ W^- \mu^+  h \mu^+ $ & $1.22\rm{e}-3$ & $1.39\rm{e}-1$ & $1.40\rm{e}+1$
& $\mu^{+} \mu^{+} \mu^{-} \nu hadr$ & - & - & 1.5 \\ 
\hline
\multicolumn{5}{|c|}{Total Cross Sections $\mu^{+} \mu^{+} \mu^{-}$ + jets + missing $E_T$} & $46.5$ & $41.7$ & $27.9$\\ 
\hline
\multicolumn{5}{|c|}{\textcolor{blue}{Total Cross Sections $\mu^{+} \mu^{+} \mu^{-}$ + jets + missing $E_T$ (only via W)}} & $36.5$ & $31.3$ & $20.8$ \\ 
\hline
\multicolumn{5}{|c|}{Total Cross Sections $\mu^{+} \mu^{+} \mu^{-}$ + missing $E_T$} & $55.1$ & $27.3$ & $13.9$ \\ 
\hline 
\multicolumn{5}{|c|}{\textcolor{blue}{Total Cross Sections $\mu^{+} \mu^{+} \mu^{-}$ + missing $E_T$ (only via W)}} & $52.8$ & $24.7$ & $12.3$ \\ 
\hline 
\hline
\multicolumn{8}{|c|}{\textcolor{blue}{Final State $+ - -$}} \\
\hline
 $ W^- \mu^+  W^- \nu $  & $8.96\rm{e}+2$ & $3.13\rm{e}+2$ & $1.39\rm{e}+2$
& $\mu^{-} \mu^{-} \mu^{+} \nu \nu \nu$ & 11.2 & 3.9 & 1.7 \\ 
\hline
 $ W^+ \mu^-  W^- \nu $  & $8.94\rm{e}+2$ & $3.11\rm{e}+2$ & $1.39\rm{e}+2$
& $\mu^{-} \mu^{-} \mu^{+} \nu \nu \nu$ & 11.1 & 3.9 & 1.7 \\ 
\hline
 $ W^- \mu^+ Z \mu^- $ & $1.27\rm{e}+2$ & $1.04\rm{e}+2$ & $5.67\rm{e}+1$  
& $\mu^{-} \mu^{-} \mu^{+} \nu hadr$ & 9.8 & 8.0 & 4.4 \\ 
\hline
 &   &  &  
& $\mu^{-} \mu^{-} \mu^{+} \nu \nu \nu $ & 2.9 & 2.4 & 1.3 \\ 
\hline
 $ W^+ \mu^- Z \mu^- $ & $1.27\rm{e}+2$ & $1.04\rm{e}+2$ & $5.67\rm{e}+1$
& $\mu^{-} \mu^{-} \mu^{+} \nu hadr$ & 9.8 & 8.0 & 4.4 \\ 
\hline
   &   &  &  
& $\mu^{-} \mu^{-} \mu^{+} \nu \nu \nu $ & 2.9 & 2.4 & 1.3 \\ 
\hline
 $ W^- \nu Z \nu $ & $2.49\rm{e}+2$ & $2.07\rm{e}+2$ & $1.13\rm{e}+2$
& $\mu^{-} \mu^{-} \mu^{+} \nu \nu \nu $ & 1.0 & 0.8 & 0.4 \\ 
\hline
 $ Z \mu^- Z \nu $ & $3.53\rm{e}+1$ & $6.93\rm{e}+1$ & $4.65\rm{e}+1$ 
& $\mu^{-} \mu^{-} \mu^{+} \nu hadr $ & 0.85 & 1.7 & 1.1 \\ 
\hline
   &   & &  
& $\mu^{-} \mu^{-} \mu^{+} \nu \nu \nu $ & 0.25 & 0.5 & 0.3 \\ 
\hline
 $ Z \mu^- h \nu $  & $3.27\rm{e}-4$ &$7.87\rm{e}-2$ & $1.12\rm{e}+1$ 
& $\mu^{-} \mu^{-} \mu^{+} \nu hadr $ & - & - & $0.37$ \\ 
\hline
 $ W^+ \nu Z \mu^- $ & $3.62\rm{e}+2$  & $3.07\rm{e}+2$ & $1.72\rm{e}+2$ 
& $\mu^{-} \mu^{-} \mu^{+} \nu hadr$ & $8.4$ & $7.2$ & $4.0$ \\ 
\hline
$ W^- \mu^+ h \mu^- $ & $5.87\rm{e}-4$ & $7.13\rm{e}-2$ & $6.86$
& $\mu^{-} \mu^{-} \mu^{+} \nu hadr$ & - & - & $0.7$ \\ 
\hline
 $ W^+ \mu^-  h \mu^- $ &$5.86\rm{e}-4$ & $7.10\rm{e}-2$& $6.87$
& $\mu^{-} \mu^{-} \mu^{+} \nu hadr$ & - & - & 0.7 \\ 
\hline
\multicolumn{5}{|c|}{Total Cross Sections $\mu^{+} \mu^{-} \mu^{-}$ + jets + missing $E_T$} & $28.9$ & $24.9$ & $15.7$ \\ 
\hline
\multicolumn{5}{|c|}{\textcolor{blue}{Total Cross Sections $\mu^{+} \mu^{-} \mu^{-}$ + jets + missing $E_T$ (only via W)}} & $19.6$ & $16.0$ & $10.2$ \\ 
\hline
\multicolumn{5}{|c|}{Total Cross Sections $\mu^{+} \mu^{-} \mu^{-}$ + missing $E_T$} & $29.4$ & $13.9$ & $6.7$ \\ 
\hline 
\multicolumn{5}{|c|}{\textcolor{blue}{Total Cross Sections $\mu^{+} \mu^{-} \mu^{-}$ + missing $E_T$ (only via W)}} & $28.1$ & $12.6$ & $6.0$ \\ 
\hline 
\end{tabular}
\end{center}
\caption{Final states with three muons for $V_e=V_\tau=0$,
  $V_\mu=0.063$. The final cross sections have been computed using the
  measured branching ratios, except for the Higgs, whose branching
  ratios have been calculated assuming a mass of 120~GeV. Only
  channels with a final cross section higher than 0.1 have been
  reported. As for the total cross sections, we have isolated the ones
  where the muons are generated via $W$ decay, since almost all the
  muons generated via $Z$ decay will be removed by the cut implemented
  to reduce the $Z$ background.}
\label{tab:trileptons++-}
\end{table}

\end{document}